\RequirePackage{lineno}
\documentclass[twocolumn,prd,nofootinbib,showpacs,preprintnumbers]{revtex4}
\usepackage{graphicx}
\usepackage{color}
\usepackage{amsmath,amssymb,amsfonts,dsfont}
\usepackage{multirow}
\usepackage{sistyle}
\usepackage{ulem}

\pdfoutput=1 

\newcommand{\lsim}{\mathrel{\mathop{\kern 0pt \rlap
  {\raise.2ex\hbox{$<$}}}
  \lower.9ex\hbox{\kern-.190em $\sim$}}}
\newcommand{\gsim}{\mathrel{\mathop{\kern 0pt \rlap
  {\raise.2ex\hbox{$>$}}}
  \lower.9ex\hbox{\kern-.190em $\sim$}}}

\newcommand{\sigmav}{$\langle\sigma v\rangle$ }

\def \aap  {A\&A}

\def \apjs  {ApJS}
\def \apjl  {ApJL}

\begin{document}


\title{The characteristics of the Galactic center excess measured with 11 years of {\it Fermi}-LAT data}

\author{Mattia Di Mauro}\email{dimauro.mattia@gmail.com}
\affiliation{Istituto Nazionale di Fisica Nucleare, via P. Giuria, 1, 10125 Torino, Italy}

\begin{abstract}
The excess of $\gamma$ rays in the data measured by the {\it Fermi} Large Area Telescope from the Galactic center region is one of the most intriguing mysteries in Astroparticle Physics.
This ``Galactic center excess'' (GCE), has been measured with respect to different interstellar emission models (IEMs), source catalogs, data selections and techniques.
Although several proposed interpretations have appeared in the literature, there are not firm conclusions as to its origin.
The main difficulty in solving this puzzle lies in modeling a region of such complexity and thus precisely measuring the characteristics of the GCE.

In this paper, we use 11 years of {\it Fermi}-LAT data, state of the art IEMs, and the newest 4FGL source catalog to provide precise measurements of the energy spectrum, spatial morphology, position, and sphericity of the GCE. 
We find that the GCE has a spectrum which is peaked at a few GeV and is well fit with a log-parabola.
The normalization of the spectrum changes by roughly 60\% when using different IEMs, data selections and analysis techniques.
The spatial distribution of the GCE is compatible with a dark matter (DM) template produced with a generalized Navarro-Frenk-White density profile with slope $\gamma = 1.2-1.3$. 
No energy evolution is measured for the GCE morphology between $0.6-30$ GeV at a level larger than $10\%$ of the $\gamma$ average value, which is 1.25.
The analysis of the GCE modeled with a DM template divided into quadrants shows that the spectrum and spatial morphology of the GCE is similar in different regions around the Galactic center.
Finally, the GCE centroid is compatible with the Galactic center, with best-fit position between $l=[-0.3^{\circ},0.0^{\circ}],b=[-0.1^{\circ},0.0^{\circ}]$, and it is compatible with a spherical symmetric morphology.
In particular, fitting the DM spatial profile with an ellipsoid gives a major-to-minor axis ratio (aligned along the Galactic plane) between 0.8-1.2.
\end{abstract}

\maketitle

\section{Introduction}
\label{sec:intro}

Several groups have reported an excess of $\gamma$ rays in the data collected by the {\it Fermi} Large Area Telescope ({\it Fermi}-LAT) in the direction of the Galactic center region, known as the Galactic center excess \citep[GCE, e,.g.][]
{Goodenough:2009gk,Hooper:2010mq,Boyarsky:2010dr,Hooper:2011ti,Abazajian:2012pn,Gordon:2013vta,Abazajian:2014fta,Daylan:2014rsa,Calore:2014xka,TheFermi-LAT:2015kwa,TheFermi-LAT:2017vmf}.
This GCE has been significantly detected when modeling the $\gamma$-ray sky using a range of interstellar emission models (IEMs), source catalogs, data selections and techniques.

Most of the references cited above agree on the fact that the GCE spectral energy distribution (SED, $E^2 dN/dE$) is peaked at few GeV.
However, other properties are much more uncertain and still under debate.
For example, the GCE has been claimed in \cite{Daylan:2014rsa} to be spherically symmetric and centered around the Galactic center.
Moreover, the GCE energy spectrum can be well modeled as $\gamma$ rays produced by dark matter (DM) particles annihilating into $b\bar{b}$ with a thermal annihilation cross section, which is the proper cross section to explain the observed density of DM in the Universe \cite{Aghanim:2018eyx}. 
All these characteristics make the GCE very appealing for indirect DM searches.

In other publications (e.g., Refs.~\cite{Calore:2014xka,TheFermi-LAT:2015kwa,TheFermi-LAT:2017vmf}) it has been shown that, using different IEMs and source catalogs, the measurement of some GCE properties, in particular the normalization of the spectrum, are too uncertain to conclude that it is of DM origin. 
For example, Refs.~\cite{Macias:2016nev,Bartels:2017vsx} claimed that the spatial shape of the GCE is better modeled with the distribution of stars in the bulge of our Galaxy. 
In particular they demonstrated that fitting the GCE with two emission templates associated with the galactic bulge provides a much better fit than using a DM template. 
This result implies that the GCE is not spherically symmetric since the boxy bulge has an ellipsoidal shape.

Refs.~\cite{Bartels:2015aea,Lee:2015fea} likewise raise doubts on a DM origin of the GCE.
By applying wavelet analysis and non-Poissonian template fitting to {\it Fermi}-LAT data, they published compelling evidence for the existence of a faint population of sources located in the Galactic center with properties that can explain the GCE.
The presence of these sources could be interpreted as a population of millisecond pulsars located around the bulge of our Galaxy.
However, very recently, Refs.~\cite{Leane:2019uhc,Chang:2019ars} have shown that the non-Poissonian template fitting method can misattribute un-modeled point sources or imperfections in the IEM to a signal of a faint population of sources or DM.
In Ref.~\cite{Buschmann:2020adf} they rerun the non-Poissonian template fitting with an improved description of the Galactic diffuse emission and found that this method continues to robustly favor the interpretation that the GCE is due, in part, to unresolved astrophysical point sources. Finally, Ref.~\cite{List:2020mzd} uses Bayesian graph convolutional neural networks and reports that the GCE is almost entirely attributed to smooth emission.
All these results challenge the robustness of the results presented in \cite{Lee:2015fea} and the conclusion that the GCE is due to a population of pulsars. 
Correspondingly, Ref.~\cite{Zhong:2019ycb} has applied wavelet analysis, similarly to what has been done in \cite{Bartels:2015aea}, to about 10 years of LAT data using the latest 4FGL catalog released by the {\it Fermi}-LAT Collaboration \cite{Fermi-LAT:2019yla}. They find that the GCE is still present, although they do not find any compelling evidence for the existence of a faint population of un-modeled sources. 

Alternative interpretations for the GCE are associated to $\gamma$ rays produced, through inverse Compton scattering, bremsstrahlung or $\pi^0$ decays, by cosmic rays injected from the Galactic center during recent outbursts \cite{Carlson:2014cwa,Petrovic:2014uda,Gaggero:2015nsa}.
These mechanisms, however, provide $\gamma$-ray signals not fully compatible with the GCE properties.
Specifically, the hadronic scenario (i.e., mostly cosmic-ray protons) predicts a $\gamma$-ray signal that is distributed along the Galactic plane, since the $\pi^0$ decay process is correlated with the distribution of gas present in the Milky Way disk \cite{Petrovic:2014uda}.  
On the other hand, a leptonic outburst would lead to a signal that is approximatively spherically symmetric.
However, this model requires a complicated scenario with at least two outbursts
to explain the morphology and the intensity of the excess with the older
outbursts injecting more-energetic electrons. 

The puzzle of the GCE origin is thus far from being solved.
Indeed, some of the GCE properties, for example its spectrum and spatial morphology, are still under debate.
Moreover, the choice of the IEM significantly changes the characteristics of the GCE, making it very challenging at the moment to make significant progress in the interpretation.
In Ref.~\cite{Dimaurosim} we have studied this problem using simulated {\it Fermi}-LAT data of the Galactic center region.
We have shown that the systematic uncertainties due to the IEM are at the level of 15\% in the normalization of the energy spectrum and 5\% in the value of the DM density profile slope used to model the GCE spatial morphology. 
Finally, we have found that an additional component in the Galactic center with a flux of the order of the {\it Fermi}-LAT bubbles can modify the results for the GCE spectrum and spatial extension at roughly the $20\%$ level.

In order to make progress on this problem, we analyze $11$ years of {\it Fermi}-LAT data using the state of the art source catalog and IEMs and new analysis techniques and we provide accurate measurements for several properties of the GCE.
We list below the main novel results with respect to previous papers on the same subject.
We use a new analysis method called weighted likelihood, recently developed by the {\it Fermi}-LAT Collaboration to mitigate the systematic uncertainties due to the mismodeling of the IEM in the Galactic plane and at energies lower than 1 GeV.  As we have demonstrated in Ref.~\cite{Dimaurosim}, this technique yields more robust results for the GCE at low energies\footnote{In order to reduce significantly the systematic uncertainties due to the modeling of the IEM improved models should be studied and used in the analysis. However, as we will explain in Sec.~\ref{sec:analysistec}, we use the weighted maximum likelihood technique that effectively reduces the differences in the results obtained with different IEMs. See Ref.~\cite{Fermi-LAT:2019yla} for a detailed discussion of this method.}.
We employ the newest 4FGL catalog of sources that contains almost twice the number of sources with respect to previous {\it Fermi}-LAT catalogs used for previous analyses.
We utilize the recently created event class {\tt SOURCEVETO} which increases the effective area of the data selections while minimizing the CR background contamination.
We make a systematic study of several aspects of the GCE such as the flux between 0.1 GeV to 1 TeV, its position, sphericity and the spatial morphology employing the state of the art IEMs provided in \cite{TheFermi-LAT:2017vmf}.
In order to derive the spatial morphology we use specific models with DM or leptonic outbursts. We also employ a model independent technique (i.e., with no dependence on the specific GCE physics interpretation) that measures, for the first time, the GCE surface brightness (i.e., the flux as a function of the angular distance) from $0^{\circ}$ up to $15^{\circ}$ and the energy dependence of its spatial morphology.
Utilizing these new techniques and analyses, we find new clues to the origin of the GCE.

The paper is organized as follow: in Sec.~\ref{sec:analysis} we describe the selection we apply to {\it Fermi}-LAT data, the choice of IEM and the analysis technique.
In Sec.~\ref{sec:refinement} we inspect how many and which components of the IEMs should be left free to provide a good fit to the data; 
in Sec.~\ref{sec:spectrum} we report our measurements for the spectrum of the GCE;
in Sec.~\ref{sec:spacial} we show the results we obtain for the spatial morphology with a model independent and with a model dependent technique;
in Sec.~\ref{sec:quadrants} we study the GCE with a DM template divided into quadrants in order to study the spatial symmetry of the excess;
and in Sec.~\ref{sec:position} we measure the GCE position relative to the Galactic center and we test its sphericity.


\section{Analysis Framework}
\label{sec:analysis} 
In this section we provide the details of the {\it Fermi}-LAT data selection, background models and analysis pipeline.

\subsection{Data selection}
\label{sec:analysisdata} 

We use 11 years of {\it Fermi}-LAT data from 2008 August 4 to 2019 August 4 passing standard data quality selection criteria\footnote{\url{https://fermi.gsfc.nasa.gov/ssc/data/analysis/documentation/Cicerone/Cicerone_Data_Exploration/Data_preparation.html}}.
We select an energy range between $0.1-1000$ GeV to study the GCE flux and spatial morphology. Then, since the GCE is significantly detected between $1-10$ GeV we focus on this energy range for analyzing its position and sphericity.
We consider a region of interest (ROI) of $40^{\circ} \times 40^{\circ}$ centered at the Galactic center. 
The size of the ROI is motivated by the extension of the GCE that can be measured significantly up to $\sim10^{\circ}$ (e.g., \cite{Calore:2014xka,TheFermi-LAT:2017vmf}).
However, we also run the analysis selecting data from a $30^{\circ} \times 30^{\circ}$ region, in order to test how the results depend on the size of the ROI.

We select photon data belonging to the Pass~8 {\tt SOURCEVETO} event class and the corresponding instrument response functions {\tt P8R3\_SOURCEVETO\_V2}.
{\tt SOURCEVETO} has the same background rate of the {\tt SOURCE} class up to 10 GeV. Above 50 GeV it is the same as the {\tt ULTRACLEANVETO}, while having 15\% more acceptance\footnote{\url{https://fermi.gsfc.nasa.gov/ssc/data/analysis/documentation/Cicerone/Cicerone_Data/LAT_DP.html}.}.
This data selection is thus ideal to analyze diffuse emission such as the GCE.
We will also show results derived with {\tt SOURCE} and {\tt ULTRACLEANVETO} data and instrument response functions to show how the GCE properties change with a different selection of the data.
The data is binned using 8 energy bins per decade and a pixel size of $0.08^{\circ}$.

\subsection{Model components}
\label{sec:analysisiem} 

The data measured in the direction of the Galactic center are modeled with the fluxes from individual sources, isotropic and IEM templates, and DM emission.
Point-like and extended sources are taken from the 4FGL {\it Fermi}-LAT catalog \cite{Fermi-LAT:2019yla} selecting all those that are in a region $48^{\circ}\times48^{\circ}$ centered at the Galactic center.
The source Sagittarius A$^{\star}$, that is usually considered to be the dynamical center of the Galaxy, is included in our background model with the source 4FGL J1745.6-2859. For a recent analysis where the spectrum, position of this source is analyzed and its cosmic-ray production is inspected see \cite{sagprep}.
We include sources up to $4^{\circ}$ outside of the ROI in order to account for the point spread function of the detector. 
The isotropic template is taken from \cite{TheFermi-LAT:2017vmf} and we leave free in the fit its normalization.
We include the $\gamma$-ray emission from the Sun, the Moon, and Loop I as in \cite{TheFermi-LAT:2017vmf}.

The spatial morphology of $\gamma$ rays produced from DM particle interactions is modeled with a DM density profile ($\rho$) parametrized using a generalized Navarro-Frenk-White (NFW) model \cite{1997ApJ...490..493N}:
\begin{equation}
\rho = \frac{\rho_0}{\left( \frac{r}{r_s} \right)^{\gamma} \left( 1 + \frac{r}{r_s} \right)^{3-\gamma}},
\label{eq:NFW}
\end{equation}
where $\rho_0$ is a normalization, $r_s$ is the scaling radius and $\gamma$ is the slope of the DM density profile.
The analysis of the Galactic center region is insensitive to the value of the scaling radius $r_s$, which we fix to 20 kpc. 
Instead, the value of $\gamma$ will be inferred from the analysis by fitting the GCE.
The $\gamma$-ray flux produced by DM annihilation is usually calculated as (see, e.g., \cite{1990NuPhB.346..129B}):
\begin{equation}
\frac{dN}{dE} (E) = \frac{1}{4\pi} \frac{\langle\sigma v\rangle}{m_\chi^2} \frac{dN_\gamma}{d E} \frac{1}{2}\mathcal{J}, 
\label{eq:flux_gamma}
\end{equation}
where the factor $1/2$ comes from considering self-conjugate DM particles, $m_\chi$ is the DM particle mass, \sigmav defines the annihilation cross section times the relative
velocity, averaged over the Galactic velocity distribution function, and $dN_\gamma/d E$ is the $\gamma$-ray production source spectrum per DM annihilation event which depends on the elementary processes ruling the annihilation of DM particles. $\mathcal{J}$ is the geometrical factor and is calculated with the integral performed in the solid angle $\Delta \Omega$ and along the line of sight (l.o.s., $s$) of the squared DM density distribution (see Eq.~\ref{eq:NFW}):
\begin{equation}
\mathcal{J} = \int_{\Delta \Omega} d\Omega \int_{l.o.s.} \rho^2 ds.
\label{eq:geomfactor}
\end{equation}
Finally, we place the center of the DM template at $(l,b)=(0^{\circ},0^{\circ})$ unless otherwise differently stated.

In this paper we are not interested in a specific DM particle physics theory interpretation of the GCE, and we accordingly
adopt a phenomenological GCE spectrum given by a log-parabola (LP)\footnote{See this page for the exact definition of this SED shape \url{https://fermi.gsfc.nasa.gov/ssc/data/analysis/scitools/source_models.html#spectralModels}.}:
\begin{equation}
\frac{dN_{{\rm{LP}}}}{dE} (E) = N_0 \left( \frac{E}{E_0} \right)^{-\alpha -\beta \log{(E/E_0)}},
\label{eq:logparabola}
\end{equation}
with the normalization $N_0$, spectral index $\alpha$, and curvature index $\beta$ free to vary in the fit. The pivot energy $E_0$ is fixed to 1 GeV.
As we will demonstrate in Sec.~\ref{sec:spectrum}, this SED shape reasonably fits the GCE spectrum.
The choice of a log-parabola model does not affect our results for the estimation of the GCE SED as for that analysis the spectral normalization is free to vary independently in each energy bin and the initial shape of the model is not important.
Moreover, when analyzing the spatial morphology, position and sphericity we select energies between 1--10 GeV and a slight change in the SED shape does not affect the results.

When we model the GCE with a DM template, we calculate its spatial morphology with the geometrical factor $\mathcal{J}$ evaluated at different longitudes and latitudes and normalized to the solid angle of the ROI ($\Delta \Omega$): $\bar{\mathcal{J}} = \mathcal{J}/\Delta \Omega$. Indeed, this is the way the {\tt Fermitools} deal with the template of extended sources\footnote{\url{https://fermi.gsfc.nasa.gov/ssc/data/analysis/scitools/extended/extended.html}}.
Finally, we compute the GCE flux with the product of a log-parabola SED (Eq.~\ref{eq:logparabola}) and the normalized geometrical factor: 
\begin{equation}
\frac{dN}{dE} = \frac{dN_{{\rm{LP}}}}{dE} \times \bar{\mathcal{J}}.
\label{eq:fluxDM}
\end{equation}

The choice of the IEM is central in modeling such a complex region as the Galactic center.
Therefore we employ the templates created in \cite{TheFermi-LAT:2017vmf} which have been optimized to model the Galactic center region.
We refer to \cite{TheFermi-LAT:2017vmf} for the full details of these models and we summarize below the main characteristics.
The templates have been created with the {\tt Galprop} code\footnote{\url{http://galprop.stanford.edu}} \cite{1998ApJ...493..694M,2000ApJ...537..763S,2004ApJ...613..962S}, which calculates the propagation and interactions of cosmic rays in the Galaxy by numerically solving the transport equations given a model for the cosmic-ray source distribution, injection spectrum, and interaction targets.
We consider 11 models generated by assuming different source distributions, interstellar gas maps\footnote{In \cite{TheFermi-LAT:2017vmf} the contribution of HI and H2 gas is considered while the one from ionized gas is neglected.}, inverse Compton components, and {\it Fermi}-LAT bubble templates.  
Therefore, these templates provide an appropriate framework, even if not exhaustive, to study how the GCE properties depend on the choice of the IEM. 
We list below and in Tab.~\ref{tab:iem} the characteristics of each IEM.

{\bf Baseline}: The reference model, labelled as {\tt Baseline}, is taken from one of the models in \cite{2012ApJ...750....3A}.
It assumes a cosmic-ray source distribution traced by the distribution of pulsars reported in \cite{Lorimer:2006qs}.
The cosmic-ray confinement volume has a height of 10 kpc and a radius of 20 kpc.
The {\tt Baseline} model assumes HI column densities derived from the 21-cm line intensities for a spin temperature of 150 K. 
The dust reddening map of \cite{1998ApJ...500..525S} is used to correct the HI maps to account for the presence of dark neutral gas not traced by the combination of HI and CO surveys \cite{2012ApJ...750....3A}. 
Moreover, it includes the inverse Compton model reported in \cite{Porter_2008} and divided into the starlight, infrared and cosmic microwave background (CMB) components. 
Finally, the model contains the Loop I, Sun, Moon emission and the {\it Fermi} bubbles divided into the low-latitude (closer to the Galactic center) and the high-latitude components.
The {\bf Baseline} model we use in this paper differs from the one used in \cite{TheFermi-LAT:2017vmf} for the inclusion of the low-latitude bubbles template.

{\bf Source distribution}: In order to account for different tracers of cosmic rays we will use different source distributions. We use an alternative model, labeled {\tt Yusifov}, generated using the pulsar distribution reported in \cite{2004AA...422..545Y}, the model {\tt SNR} created with distribution of supernova remnants published in \cite{1998ApJ...504..761C} (labelled as {\tt SNR}) and finally with the distribution of OB stars from \cite{2000AA...358..521B} ({\tt OBstars}).
Finally, we test a model labeled as {\tt Pulsars} where we include the bremsstrahlung and $\pi^0$ emission divided into the neutral atomic (HI) and molecular hydrogen (H2) components.

{\bf Gas and inverse Compton}: We test, in a model labeled {\tt ICS combined}, an inverse Compton template where the starlight, infrared and CMB components are merged in a unique template.
In order to account for different gas models we use templates generated from maps of the starlight extinction due to interstellar dust taken from the Variables in the Via Lactea survey \cite{2010NewA...15..433M} ({\tt SLext}) and using the high-resolution maps from the GASS survey \cite{2010AA...521A..17K} and the dust extinction map from extinction map from \cite{2014AA...571A..11P} that is built using IRAS and Planck data (labelled as {\tt PlanckGASS})

\textbf{\textit{Fermi}} {\bf Bubbles}: The {\it Fermi} bubbles were discovered in \cite{2010ApJ...724.1044S}
Recently, \cite{TheFermi-LAT:2017vmf,Herold:2019pei} modeled this component dividing the emission in two parts: a low-latitude and a high-latitude component. 
The low-latitude component is the one that has the largest impact on the GCE. Therefore, in the model {\tt No low-lat Bubbles} we decide to neglect the presence of this component to study how this affects the results.

\begin{figure*}[t]
\includegraphics[width=0.49\textwidth]{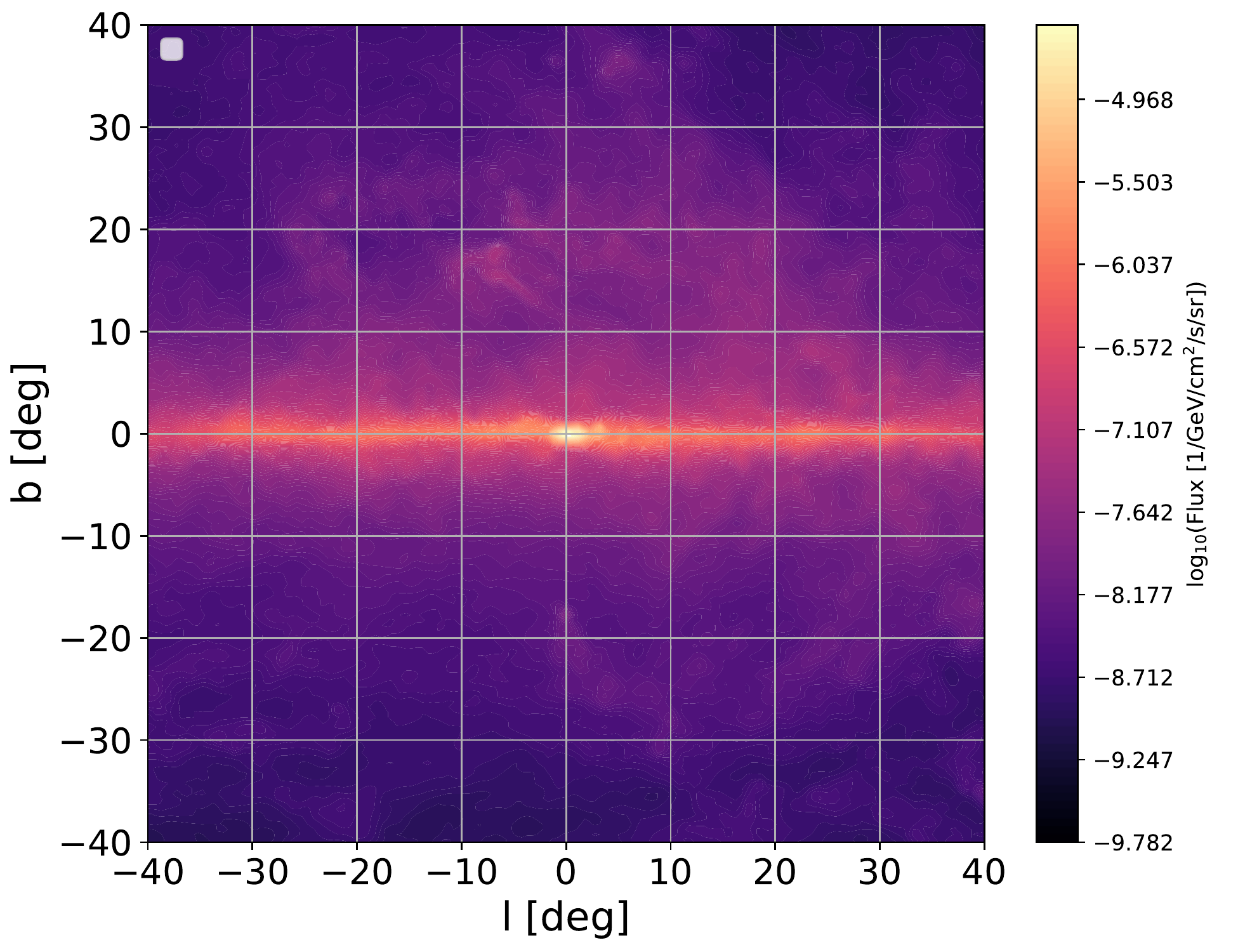}
\includegraphics[width=0.49\textwidth]{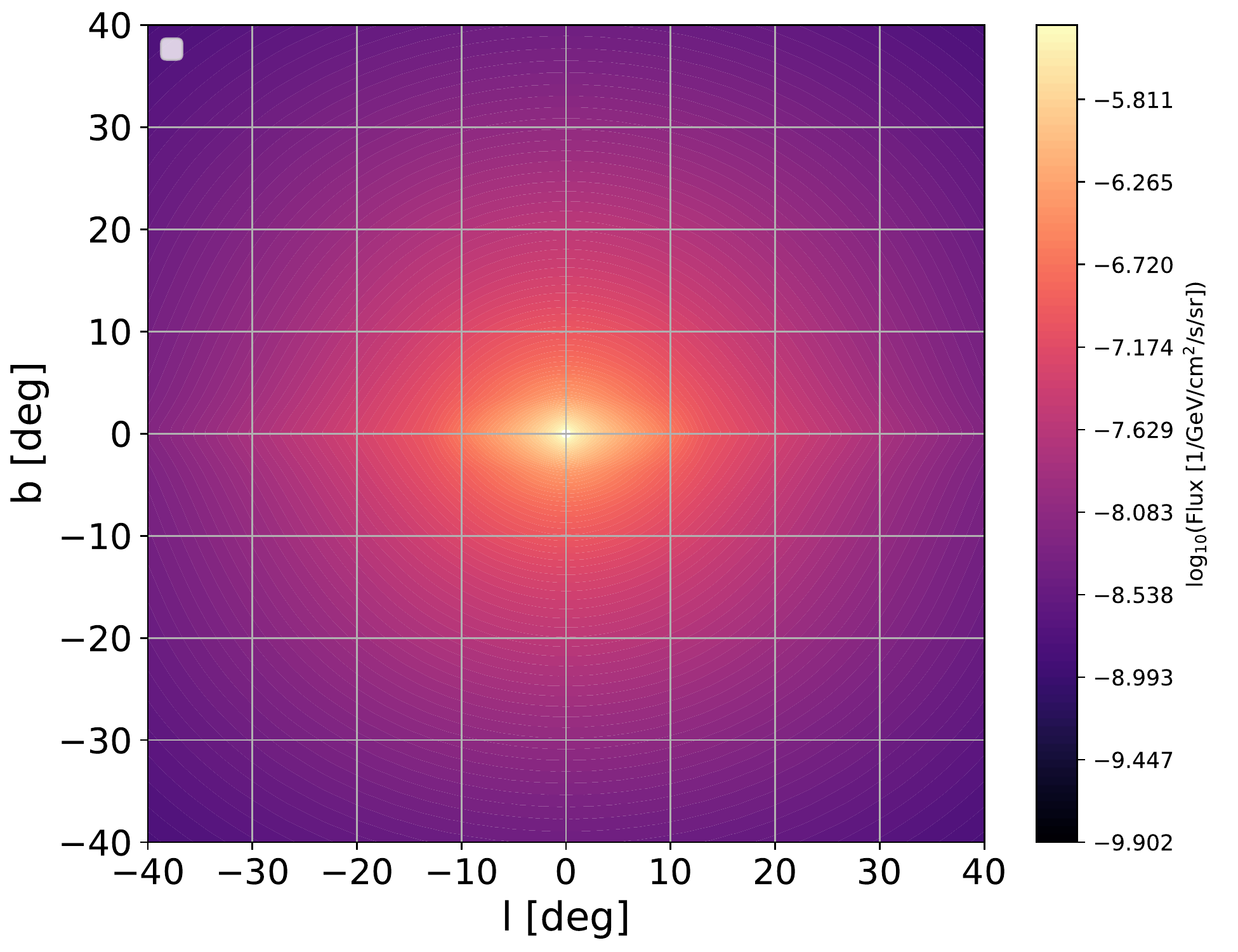}
\caption{Map of the flux calculated at 1 GeV for the $\gamma$-ray emission produced for bremsstrahlung and $\pi^0$ from cosmic-ray protons (left panel) and for inverse Compton scattering from electrons and positrons injected from the Galactic center (right panel). The color bar represents the intensity of the emission in the different directions of the ROI in units of 1/MeV/cm$^2$/s/sr. We consider the case with a vertical size of the Galactic halo of $z=8$ kpc. The maps reported here have been produced in Ref.~\cite{TheFermi-LAT:2017vmf}.}
\label{fig:fluxmap}
\end{figure*}

{\bf Additional cosmic rays in the Galactic center}: We test the presence of an additional population of cosmic rays injected from the Galactic center considering the following two processes: cosmic-ray protons injected in the central molecular zone and electrons and positrons emitted from the Galactic bulge. 
In the former, $\gamma$ rays are produced for bremsstrahlung and $\pi^0$ from protons interacting with the gas in the central molecular zone. 
In the latter, photons are generated from electrons and positrons emitted from the Galactic bulge and inverse Compton scattering on the interstellar radiation field in the inner part of the Galaxy.
These additional cosmic-ray electrons and positrons and protons are considered both with the model labeled as {\tt CMZ}.
Instead, in the model named as {\tt IC bulge} only the electrons and positrons produced from the Galactic bulge are added.
For the $\gamma$-ray emission produced from protons, the tracer of the cosmic-ray production in the central molecular zone is taken from the distribution of molecular gas in Equation 18 from \cite{2007AA...467..611F}. 
In the model with electrons and positrons the population of MSPs in the Galactic bulge is modeled assuming that their distribution is traced by the old stellar population in the bulge from \cite{2012AA...538A.106R}. 
For both these models we tried different vertical sizes $z$ of the propagation halo from 4 and 8 kpc finding for all these values similar results.
In Fig.~\ref{fig:fluxmap} we show the flux map at 1 GeV for the $\gamma$ rays produced for bremsstrahlung and $\pi^0$ from cosmic-ray protons and for inverse Compton scattering from electrons and positrons and assuming $z=8$ kpc.
The signal produced for bremsstrahlung and $\pi^0$ decay from protons follows the distribution of interstellar gas, thus the signal is elongated on the Galactic plane and far from being spherically symmetric.
The $\gamma$-ray emission from electrons and positrons traces the distribution of the interstellar radiation field in the inner part of the Galaxy and the shape of the Galactic bulge. The morphology is not spherically symmetric but rather similar to an ellipsoid with the ratio between the major and minor axis equal to about 1.7, similarly to what considered for this component in \cite{Macias:2016nev}.

{\bf Analysis setup}: In addition to the choice of different IEMs we also test our analysis for different data selections. 
We consider the {\tt SOURCE} and {\tt ULTRACLEANVETO} data and correspondent instrument response functions.
The LAT point spread function (PSF) is a function of the incident photon's energy and inclination angle, and the event class. Using an event-level quantity indicating the quality of the reconstructed direction, the data is divided into quartiles, from the lowest quality quartile (PSF0) to the best quality quartile (PSF3)\footnote{Visit the following page for further details \url{https://fermi.gsfc.nasa.gov/ssc/data/analysis/documentation/Cicerone/Cicerone_Data/LAT_DP.html}}. 
We test the selection of data for event type PSF 2 and 3, i.e., with the best reconstructed direction, and {\tt SOURCEVETO} instrument response functions ({\tt PSF23}).
Moreover, we make a test with a smaller ROI of size $30^{\circ}\times 30^{\circ}$ ({\tt ROI $30\times30$}). 
Finally, we run the analysis both with and without ({\tt no weights}) the weighted maximum likelihood analysis, as described in Sec.~\ref{sec:analysistec}.

\begin{table*}
\begin{center}
\begin{tabular}{|c|c|c|}
\hline
IEM name & Assumption tested  & Model used   \\
\hline 
{\tt Yusifov}   & Source distribution  &  Pulsar distribution from \cite{2004AA...422..545Y}  \\ 
{\tt SNR}   & Source distribution  &  SNR distribution from \cite{1998ApJ...504..761C}   \\ 
{\tt OBstars}   & Source distribution  & OB stars distribution from \cite{2000AA...358..521B}  \\ 
{\tt Pulsars}   & Source distribution  & Pulsar distribution from \cite{Lorimer:2006qs} and divided into HI, H2  \\ 
\hline
\hline
{\tt ICS combined}   & Gas and inverse Compton & unique template for the inverse Compton component  \\ 
{\tt SLext}   & Gas and inverse Compton & gas model derived from starlight extinction maps \cite{2010NewA...15..433M}  \\ 
{\tt PlanckGASS}   & Gas and inverse Compton & gas model derived from Planck and GASS data \cite{2014AA...571A..11P,2010AA...521A..17K}  \\ 
\hline
\hline
{\tt No low-lat Bubbles}   & {\it Fermi} Bubbles & no low-latitude {\it Fermi} Bubbles   \\ 
\hline
\hline
{\tt CMZ}   & Additional cosmic rays & protons injected from the CMZ and electrons from the Galactic bulge   \\ 
{\tt IC bulge}   & Additional cosmic rays & electrons injected from the Galactic bulge   \\ 
\hline
\hline
{\tt SOURCE}   & Analysis setup & {\tt SOURCE} data and instrument response function   \\ 
{\tt ULTRACLEANVETO}   & Analysis setup & {\tt ULTRACLEANVETO} data and instrument response function   \\ 
{\tt PSF23}   & Analysis setup &  {\tt SOURCEvet} data and event type PSF23 \\ 
{\tt ROI $30\times 30$}   & Analysis setup &  $30^{\circ}\times 30^{\circ}$ ROI \\ 
{\tt no weights}   & Analysis setup &  no weighted likelihood analysis \\ 
\hline
\hline
\end{tabular}
\caption{This table summarizes the IEM and analysis setup and techniques used in the paper. For each model we report the name we use throughout the paper and the assumption that we test with it. The complete description of each model is reported in Sec.~\ref{sec:analysisiem} .}
\label{tab:iem}
\end{center}
\end{table*}

\subsection{Analysis technique}
\label{sec:analysistec} 

Our analysis pipeline is entirely based on {\tt FermiPy}, a Python package that automates analyses with the {\tt Fermitools} \cite{2017ICRC...35..824W}\footnote{See \url{http://fermipy.readthedocs.io/en/latest/}.}.
{\tt FermiPy} includes tools to perform high-level analysis of {\it Fermi}-LAT data.
We employ  versions {\tt 18.0.0} of {\tt Fermipy} and {\tt 1.2.3} of the {\tt Fermitools}.


The pipeline we apply is based on the following steps.
We run the {\tt gta.setup()} tool which makes cuts on the data according to the time and energy range, spatial and energy bins considered and the instrument response functions selected.
Then, we apply the {\tt gta.optimize()} tool that fits the model to the data with an iterative strategy. 
First, it simultaneously fits the normalizations of the brightest components and sources. 
Then, it individually fits the normalizations of all sources that are not included in the first step.  
Finally, it fits the shape and normalization parameters of each source simultaneously. 
This tool is a fast method to reach a good agreement between model and data. 
In the absence of strong degeneracies between the different components, i.e.~high latitudes and selection of data above 1 GeV, it provides a model that is basically identical to the one found with a complete likelihood fit (i.e., with all the parameters free to vary at the same time).
However, since the Galactic center includes many sources and components, many degeneracies are present and we have to perform a complete fit.
This is done by running the {\tt gta.fit()} tool which is a Python wrapper of the pyLikelihood fit method implemented in the {\tt Fermitools}. 
This tool returns the best-fit SED parameters and the full covariance matrix.

After this first step, we remove from the model the sources detected with a Test Statistic ($TS$)\footnote{The Test Statistic ($TS$) is defined as twice the difference in maximum log-likelihood between the null hypothesis (i.e., no source present) and the test hypothesis: $TS = 2 ( \log\mathcal{L}_{\rm test} -\log\mathcal{L}_{\rm null} )$~\cite{1996ApJ...461..396M}.} lower than 25. This is indeed the usual cut in $TS$ that is used to include sources in {\it Fermi}-LAT catalogs. A $TS$ of 25 corresponds roughly to a detection at $5\sigma$ significance.
We then re-optimize the likelihood to obtain the final model. 
In the last step, we calculate the GCE SED using the {\tt gta.sed} command.
This tool fits the data by varying the flux normalization independently in each energy bin. It assumes a power-law SED in each energy bin with a fixed spectral index of 2.0. 
This is a good approximation with small enough bins as we have in our analysis. 
During this step of the analysis we leave free in the fit the SED parameters of the GCE, IEM templates and sources detected with a $TS>1000$ and located at a distance smaller than $10^{\circ}$ from the Galactic center. 
Therefore, the results obtained with {\tt gta.sed} are independent from the SED model assumed initially for the DM component.
Our results for the GCE SED are thus model independent and can be applied a posteriori to search possible interpretations of the GCE.

In \cite{Dimaurosim} we have tested this pipeline with simulated data of the Galactic center.
We have verified that it recovers properly the injected signal of DM (specifically its spatial morphology and flux) and the correct SED parameters of background sources.
Finally, we have shown, by simulating the data with one model and using the same model in the analysis, the pipeline does not leave significant residuals. Indeed, the $TS$ map is compatible with the $\chi^2/2$ distribution for 1 degree of freedom, as it should be in case of a perfect knowledge of the model (see the right panel of Fig.~1 in \cite{Dimaurosim}).

In order to characterize the spatial morphology of the GCE, we avoid the use of specific spatial templates and instead model the excess as a set of concentric, uniform annuli.
This method consists of including in the model concentric uniform annuli and fitting them to the data simultaneously.
The SED of each annulus is modeled with a power law, i.e., the free parameters for each annulus are the normalization and the power-law index.
Since we will apply this analysis to small energy bins the power-law approximation is appropriate.
The fitting procedure we employ is the same explained above.
After performing the fit, we extract the energy flux ($S$) of each annulus in units of MeV cm$^{-2}$ s$^{-1}$ and we compute the surface brightness of the GCE as $dN/d\Omega$, i.e., by dividing the annulus energy fluxes by their solid angles $d\Omega$.
In particular in \cite{Dimaurosim} we have demonstrated that we are able to recover the standard deviation of an injected 2D Gaussian signal and the value of $\gamma$ for an injected DM signal.

We also use a model-dependent technique to find the GCE spatial morphology by fitting the data with DM templates generated using a NFW density profile (see Eq.~\ref{eq:NFW}) with different values for $\gamma$.
The background components are the same for all the test cases. 
These likelihood values, obtained by optimizing with respect to all other parameters, yield the profile likelihood $\rm{Log} (\mathcal{L})(\gamma)$, from which we can obtain estimates of $\gamma$ and its uncertainties with standard maximum likelihood methods.
Finally, by maximizing $\rm{Log} (\mathcal{L})(\gamma)$ we find the best-fit value and errors for $\gamma$.
We have demonstrated in \cite{Dimaurosim} that, in case the GCE is simulated with a DM template, the model independent technique with the annuli and the results of the $\rm{Log} (\mathcal{L})(\gamma)$ method provide very similar results for the $\gamma$ parameter.

\begin{figure*}
\includegraphics[width=0.49\textwidth]{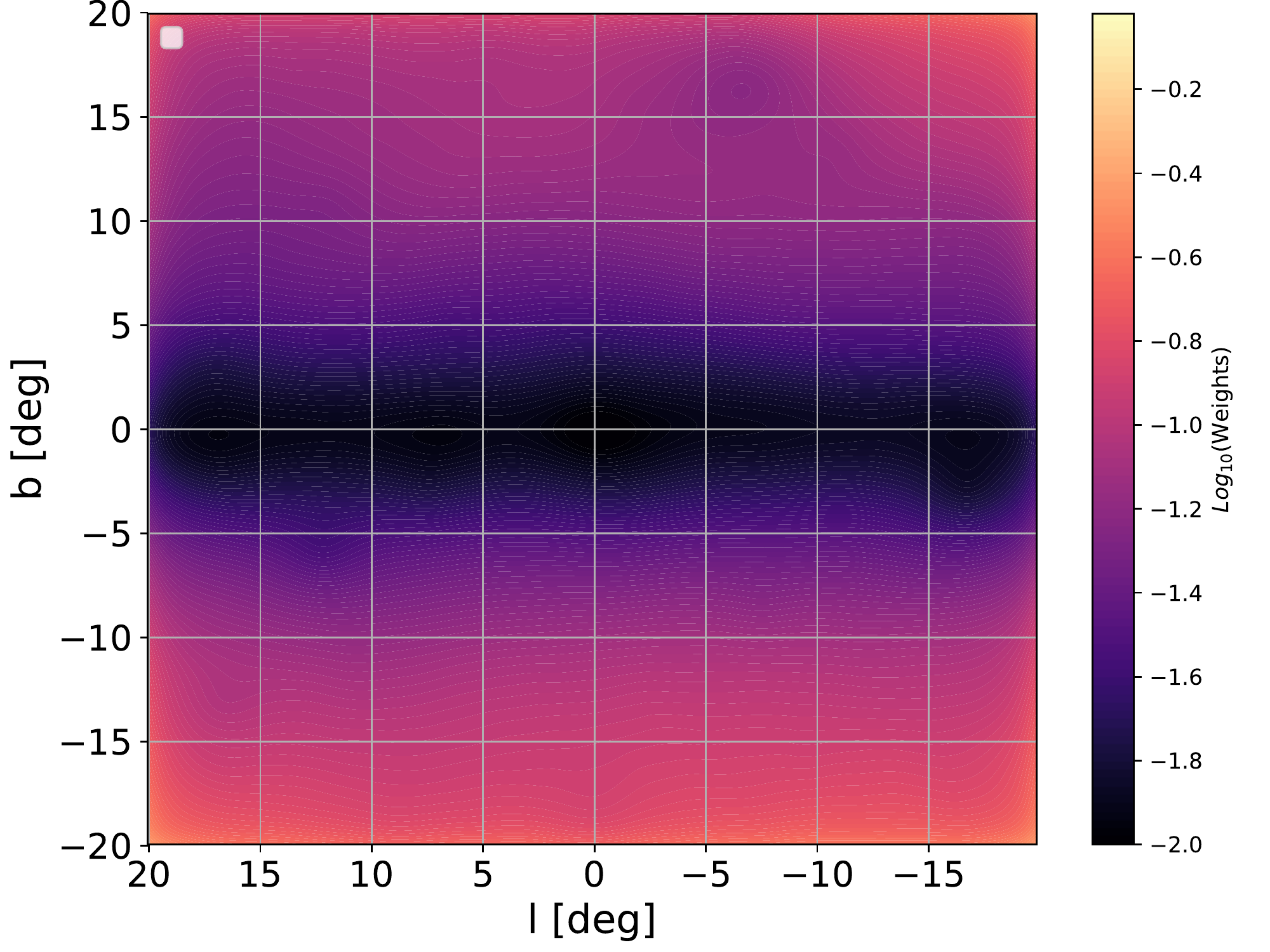}
\includegraphics[width=0.49\textwidth]{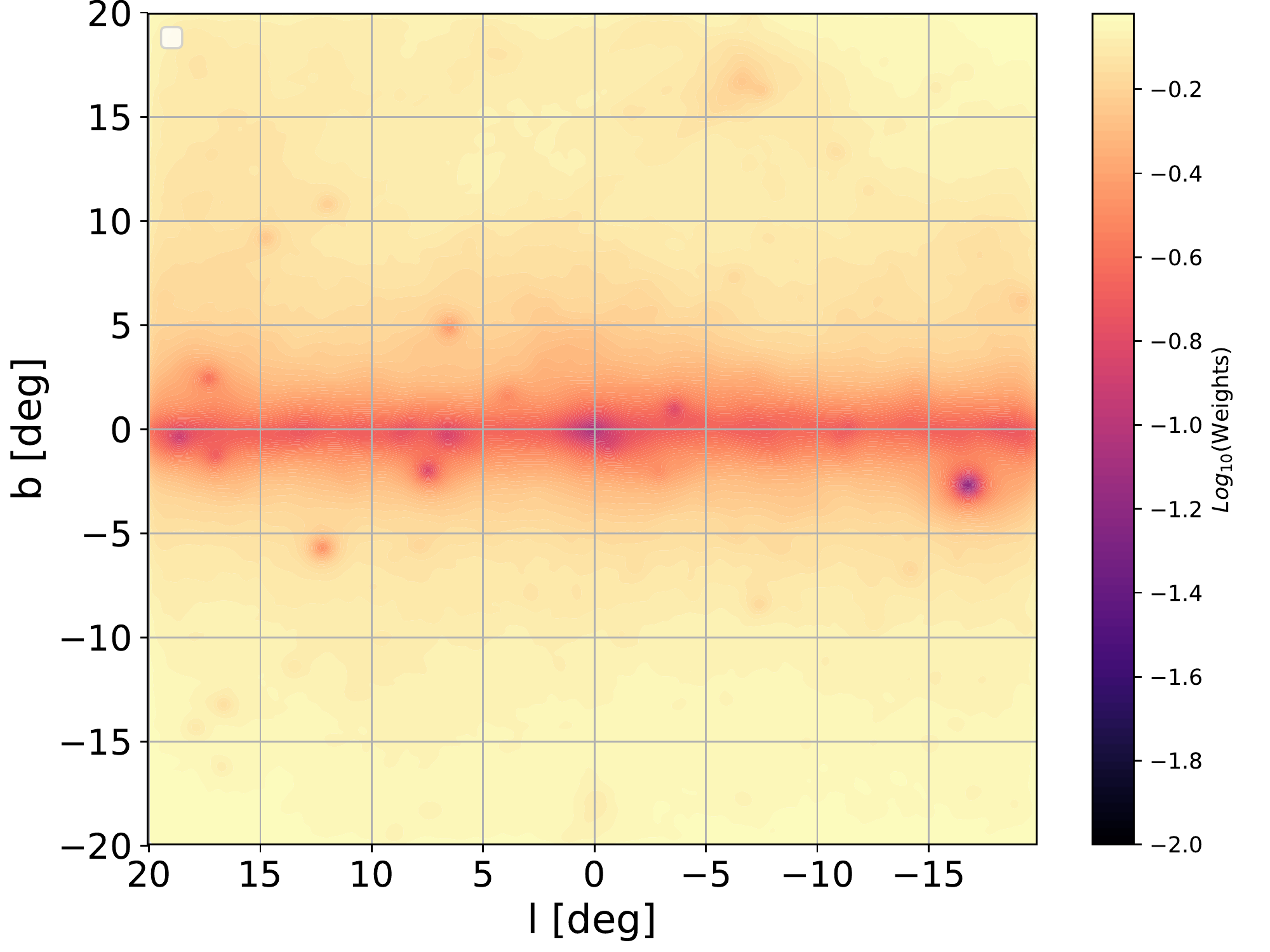}
\caption{Weight maps generated for $E=0.27$ GeV (left panel) and $1.1$ GeV (right panel). The color bar represents the $\rm{Log}_{10}$ of the weight values. See the main text for further details on the weighted likelihood technique.}
\label{fig:weights}
\end{figure*}

The region where the uncertainty of the IEM is the largest is along the Galactic plane ($b\sim0^{\circ}$).
In order to mitigate these uncertainties in our analysis, we apply the weighted likelihood technique that has been recently included in the {\tt Fermitools}.
This technique introduces weights for every pixel of the sky according to the number of counts present in the data.
These weights are then multiplied by the $\rm{Log} (\mathcal{L})$ found in each pixel \cite{Fermi-LAT:2019yla}. 
Effectively this procedure penalizes pixels with a very large number of photons in which the systematics for the choice of the IEM are larger\footnote{A technical document explaining the weighted likelihood is available at this link \url{https://fermi.gsfc.nasa.gov/ssc/data/analysis/scitools/weighted_like.pdf}.}.
We use as, in the 4FGL catalog paper, a systematic level of $\epsilon=3\%$. This value is motivated by the study performed in Ref.~\cite{Fermi-LAT:2019yla} with the relative spatial and spectral residuals in the Galactic plane where the diffuse emission is strongest. 
We show in Fig.~\ref{fig:weights} the weight maps derived at energies 0.27 GeV and 1.1 GeV in our ROI.
At low energy the weights for $|b|<3^{\circ}$ are very small. This implies that all these pixels constrain much less the fitting procedure than the higher latitude ones where the weights are much larger.
Instead, at 1.1 GeV most of the ROI has weights close to 1 meaning that most of pixels have the same weights.
The effect of the weighted likelihood is thus important below 1 GeV and in the inner few degrees from the Galactic plane where the IEMs differ the most.

Finally, we apply the energy dispersion to all the components of our model using the method implemented in the {\tt Fermitools}\footnote{For a complete description see \url{https://fermi.gsfc.nasa.gov/ssc/data/analysis/documentation/Pass8_edisp_usage.html}.}.
Specifically, we select {\tt apply\_edisp=true} and {\tt edisp\_bins=-1} which applies the energy dispersion only on the spectrum accounting for one extra bin.

\section{Results}

\subsection{Interstellar emission components}
\label{sec:refinement}
Previous papers on the GCE have made different choices for the number of IEM components to leave free in the analysis.
For example in \cite{Goodenough:2009gk,Hooper:2010mq,Hooper:2011ti,Daylan:2014rsa} the IEM was modeled with a single template, and Ref.~\cite{Daylan:2014rsa} used the IEM released with Pass 6 data. 
Instead Refs.~\cite{Calore:2014xka,TheFermi-LAT:2015kwa,TheFermi-LAT:2017vmf} divided the IEM in multiple templates generated with {\tt Galprop}. 
Specifically, Ref.~\cite{Calore:2014xka} uses 60 different IEMs, each divided into the components we reported in sec.~\ref{sec:analysisiem}.
On the other hand Refs.~\cite{TheFermi-LAT:2015kwa,TheFermi-LAT:2017vmf} consider the inverse Compton, bremsstrahlung and $\pi^0$ components divided into rings.

\begin{figure*}
\includegraphics[width=0.49\textwidth]{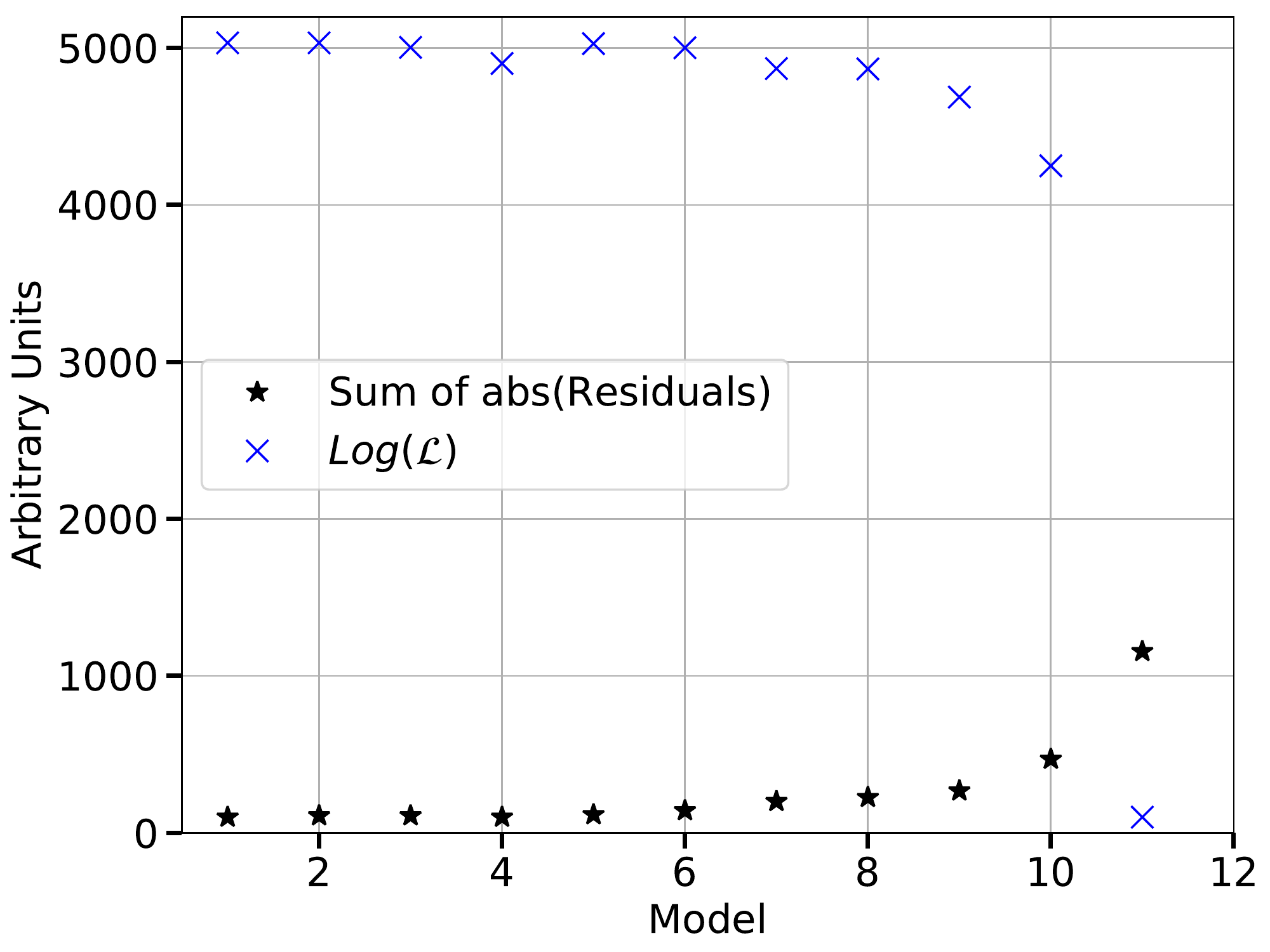}
\includegraphics[width=0.49\textwidth]{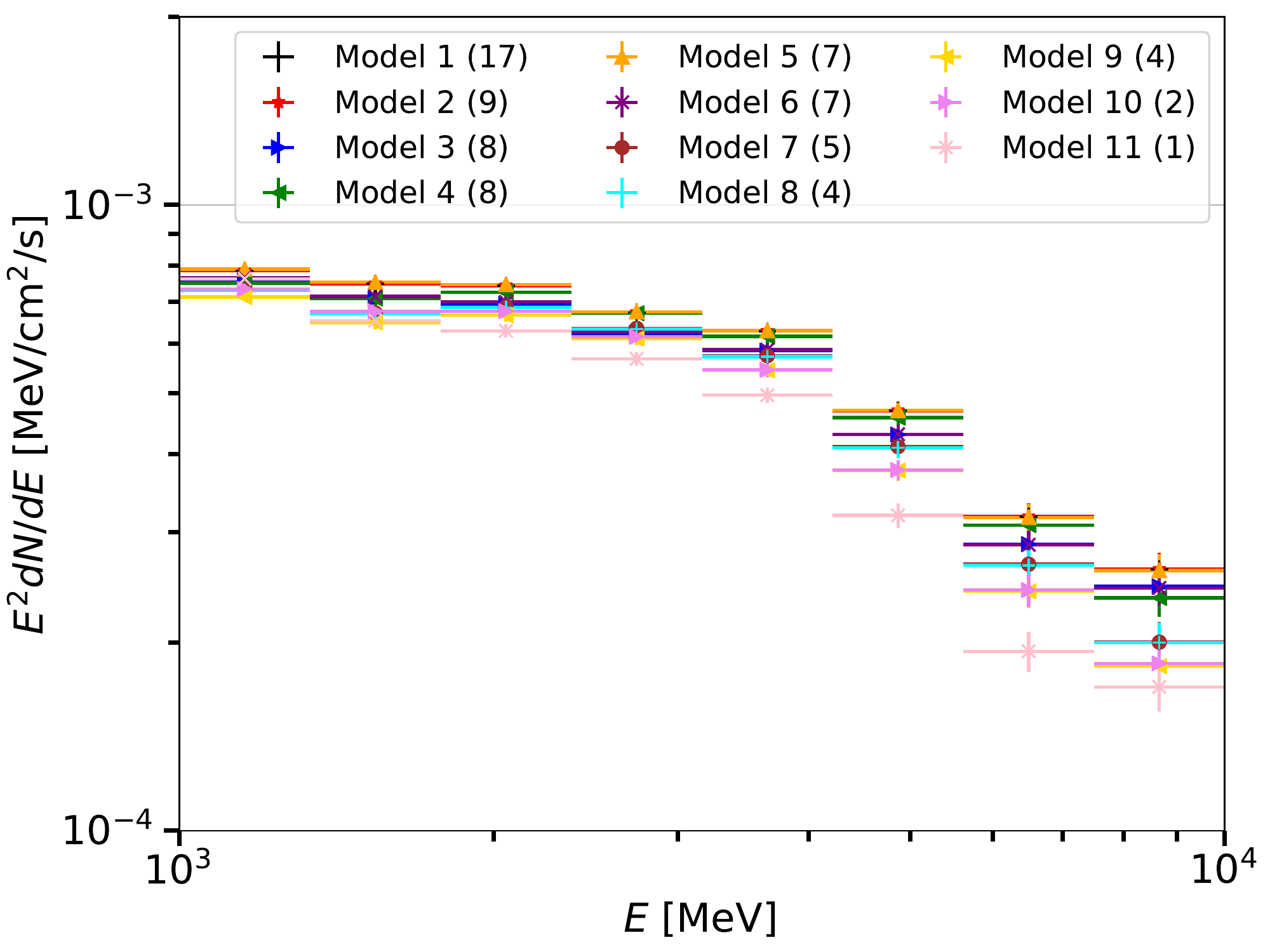}
\caption{Left Panel: value of the $\rm{Log}(\mathcal{L})$ and sum of the absolute value of the residuals obtained when we run the analysis pipeline to {\tt Model i} with $i=[1,11]$ (see Tab.~\ref{tab:models} for further details.).
Right Panel: SED of the GCE obtained with the different {\tt Models}. In parenthesis of the legend we report the number of components present for each {\tt Model}.}
\label{fig:trycomp}
\end{figure*}

\begin{figure}
\includegraphics[width=0.49\textwidth]{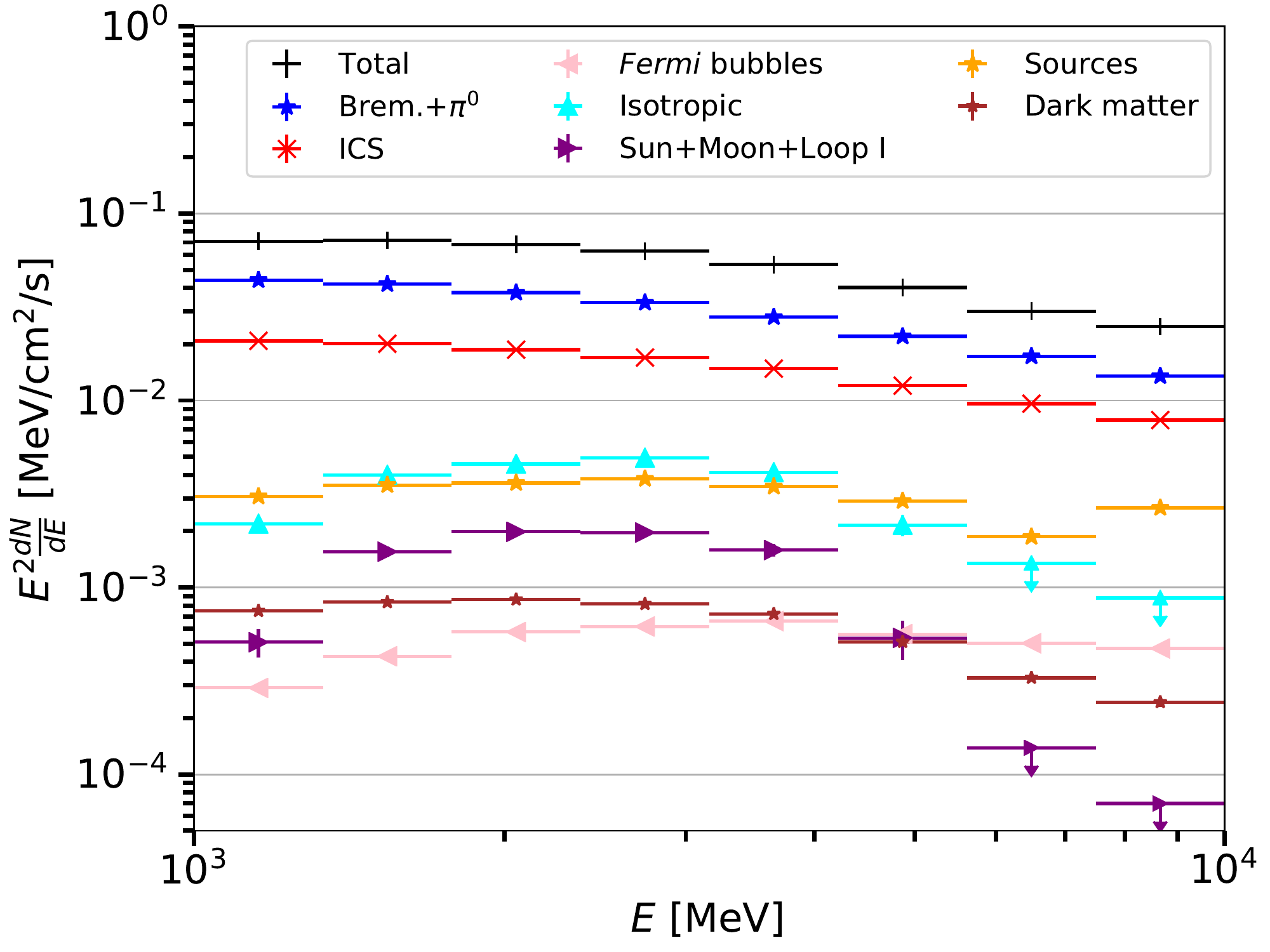}
\caption{Energy spectrum between 1 and 10 GeV of the different components included in our analysis with {\tt Model 2}. Here, we use the components generated with the {\tt Baseline} IEM (see Sec.~\ref{sec:analysisiem} for further details).}
\label{fig:spectrumcomp}
\end{figure}

\begin{figure*}[t]
\includegraphics[width=0.49\textwidth]{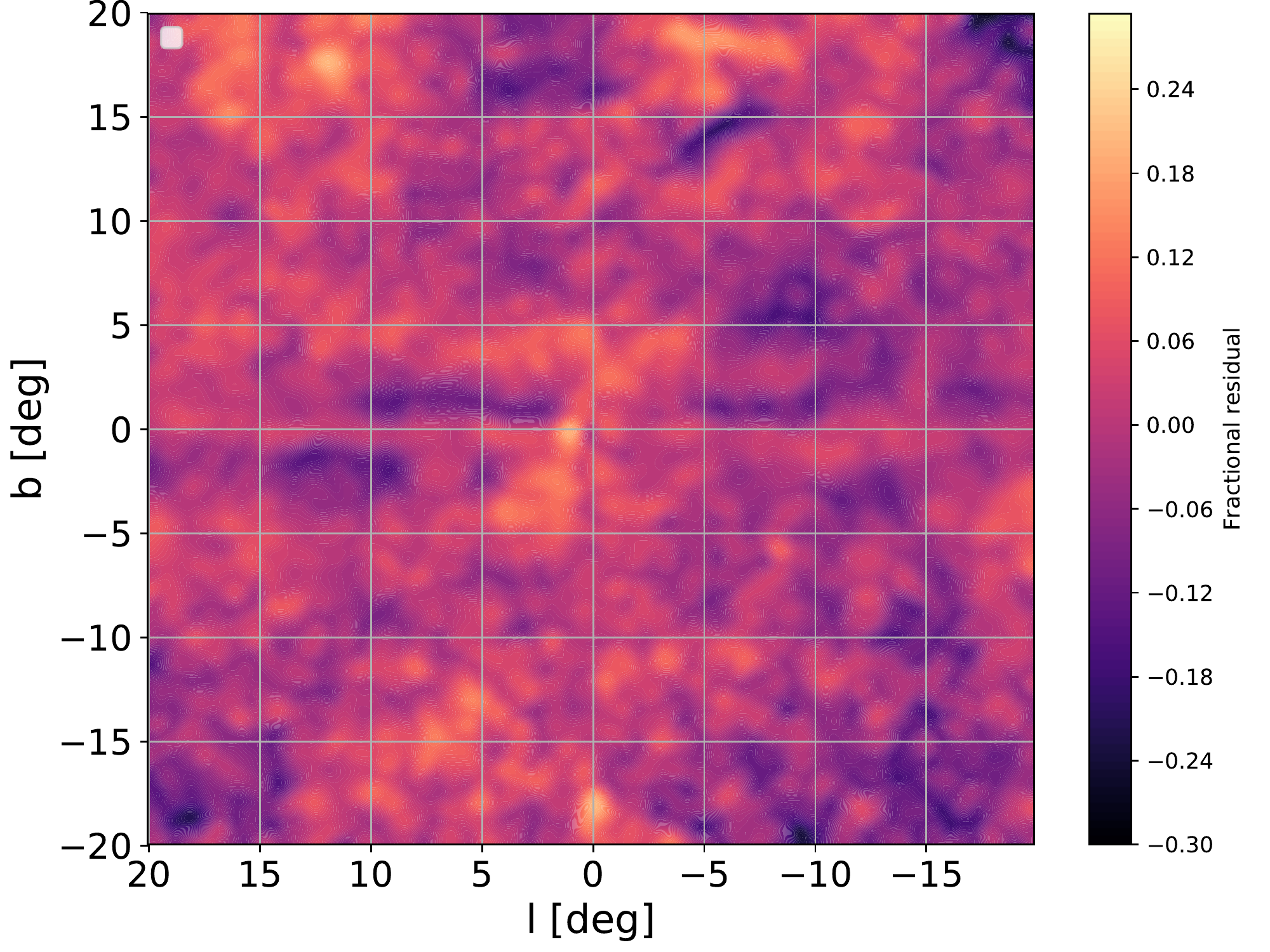}
\includegraphics[width=0.49\textwidth]{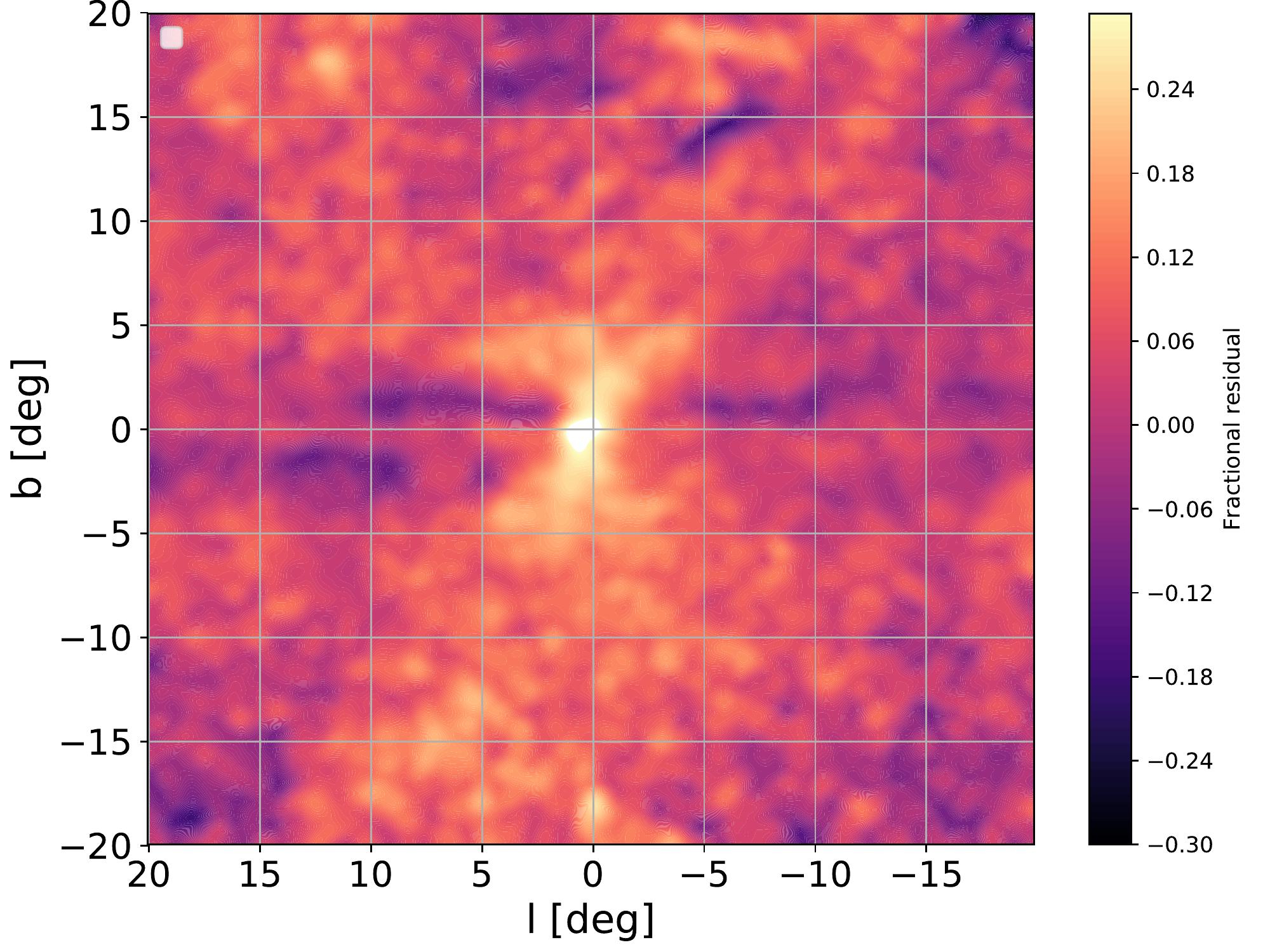}
\caption{We show in the left panel the map of the residuals obtained using the {\tt Model 2} and the {\tt Baseline} IEM to fit the ROI in the energy range 1-10 GeV. Instead, in the right panel we have added to the residuals the counts of the GCE. The different colors represent the fractional residual, i.e., the residual counts (data minus model) divided by the total counts.}
\label{fig:residmap}
\end{figure*}

In order to define the appropriate number of IEM components, we apply the following analysis.
We consider different cases labeled as {\tt Model i} with {\tt i} ranging from 1 to 11. 
Models with increasing {\tt i} include a decreasing number of IEM components.
The DM template is always separated from all the other background sources. 
We consider for all the {\tt Models} the same list of point and extended sources which has been found by using the analysis pipeline described in Sec.~\ref{sec:analysistec}.
In particular we use the templates generated with the case labeled as {\tt Baseline} in Sec.~\ref{sec:analysisiem} and we assume different choices for the number of components left free in the fit.
In Tab.~\ref{tab:models}, we report a summary of the templates that are left free for each {\tt Model}.
In the next sections we will fix the number and type of components left free in the fit and will test the different IEMs and analysis setup listed in Sec.~\ref{sec:analysisiem}.


\begin{table*}
\begin{center}
\begin{tabular}{|c|c|c|}
\hline
{\tt Model} & N$^{\circ}$ comp. & list templates used for each component   \\ 
{\tt 1}        &       17                    & 5 rings for bremsstrahlung (Brem.) and $\pi^{0}$ emission, 3 for inverse Compton (CMB, SL. IR),\\ 
                &                              & 2 for low latitude bubbles, isotropic (ISO), 1 template for Loop I, the Sun and the Moon (LoopMoonSun).   \\ 
{\tt 2}        &       9                    & 1 Brem., 1 $\pi^{0}$, 3 for inverse Compton (IC), 2 for low latitude bubbles, 1 ISO, 1 LoopMoonSun.   \\ 
{\tt 3}        &       8                    & 1 Brem. and $\pi^{0}$, 3 IC, 2 for low latitude bubbles, 1 ISO, 1 LoopMoonSun.   \\ 
{\tt 4}        &       8                    & 1 Brem., 1 for $\pi^{0}$, 3 IC, 1 bubbles, 1 ISO, 1  LoopMoonSun.   \\ 
{\tt 5}        &       8                    & 2 for Brem. and $\pi^{0}$ divided into H1 and H2, 3 IC, 1 bubbles, 1 ISO, 1 LoopMoonSun.   \\
{\tt 6}        &       7                    & 1 Brem., 1 for $\pi^{0}$, 1 for IC, 2 for bubbles, 1 ISO, 1  LoopMoonSun.   \\ 
{\tt 7}        &       7                    & 1 Brem. and $\pi^{0}$, 3 IC, 1 bubbles, 1 ISO, 1 LoopMoonSun.   \\
{\tt 8}        &       5                    & 1 Brem. and $\pi^{0}$, 1 IC, 1 bubbles, 1 ISO, 1 LoopMoonSun.   \\
{\tt 9}        &       4                    & 1 Brem. and $\pi^{0}$, 1 IC, 1 bubbles, 1 ISO and LoopMoonSun.   \\
{\tt 10}      &       4                    & 1 Brem. and $\pi^{0}$ and IC, 1 bubbles, 1 ISO, 1 LoopMoonSun.   \\
{\tt 11}      &       1                    & 1 unique template for all components.   \\
\hline
\hline
\end{tabular}
\caption{This table summarizes the templates that we leave free for each {\tt Model}.}
\label{tab:models}
\end{center}
\end{table*}

In this analysis, we select data between 1 and 10 GeV since at these energies the GCE is brighter and more significant.
We apply the analysis pipeline explained in Sec.~\ref{sec:analysistec} to all the {\tt Models} reported before.
We then save the value of the log-likelihood ($\rm{Log}(\mathcal{L})$) obtained in the end of the analysis and we generate a residual map to check how well each model fits the data.
In the left panel of Fig.~\ref{fig:trycomp} we show the value of $\rm{Log}(\mathcal{L})$ obtained for each {\tt Model}.
Moreover, we calculate the sum of the absolute values of the residuals in all the pixels of the ROI. This is shown also shown in Fig.~\ref{fig:trycomp}.
The figure demonstrates that we obtain similar $\rm{Log}(\mathcal{L})$ values and sum of the residuals if we use {\tt Model $i$} with $i<6$. 
Instead, for models with a smaller number of components ($i>6$) the $\rm{Log}(\mathcal{L})$ decreases and the residuals increase significantly (meaning that the fit becomes poorer).

We also calculate the GCE SED obtained with the different {\tt Models} and we show it in the right panel of Fig.~\ref{fig:trycomp}.
The SED obtained is similar at energies $1-2$ GeV while at higher energies {\tt Model} 6 to 11 seem to give slightly steeper SED compared to all the other cases.
The difference in the SED obtained for {\tt Model} 1 to 6 is at most 5\% which is much smaller than the differences obtained when using different IEMs as we will see in Sec.~\ref{sec:spectrum}. Therefore, the systematics we find due to the choice of the number of components to leave free in the fit (assuming {\tt Model i} with $i<6$) is negligible with respect to the systematics due to the choice of the specific IEM physics. 

We decide to use in our analysis {\tt Model 2} that provides, as well as the others with $i<6$, a good fit to the data, small residuals, has a number of components much smaller than {\tt Model 1} and about the same number of components for the ones with $2<i<6$.  {\tt Model 2} is divided into the following parts: two separate templates for the bremsstrahlung and $\pi^{0}$ emission, three templates for the inverse Compton emission, one for each interstellar radiation field (starlight, infrared and CMB), two components for the {\it Fermi} bubbles (low and high-latitude as in \cite{TheFermi-LAT:2017vmf}), the isotropic component and one unique template for the emission from the Moon, Sun and Loop I.
In Fig.~\ref{fig:spectrumcomp} we show the flux of the different components as a function of energy between 1 and 10 GeV.
The component with the largest flux is due to the bremsstrahlung and $\pi^{0}$ processes which contributes roughly 60\% of the observed flux.
The inverse Compton is at about 30\% of the total flux. The isotropic emission and the cumulative flux of all sources contribute together at most about 10\%.
The GCE flux modeled with a DM template with $\gamma=1.2$ is at about a factor of 1\% of the total flux.
The {\it Fermi} bubbles also have a similar flux.

In Fig.~\ref{fig:residmap} we show the fractional residuals, i.e., the residuals divided by the total number of counts, and the residuals plus the GCE counts obtained with {\tt Model 2} 
The residuals are of the order of $20-25\%$ of the counts, similarly to that found in \cite{Calore:2014xka,TheFermi-LAT:2015kwa,TheFermi-LAT:2017vmf}.

\begin{figure}
\includegraphics[width=0.49\textwidth]{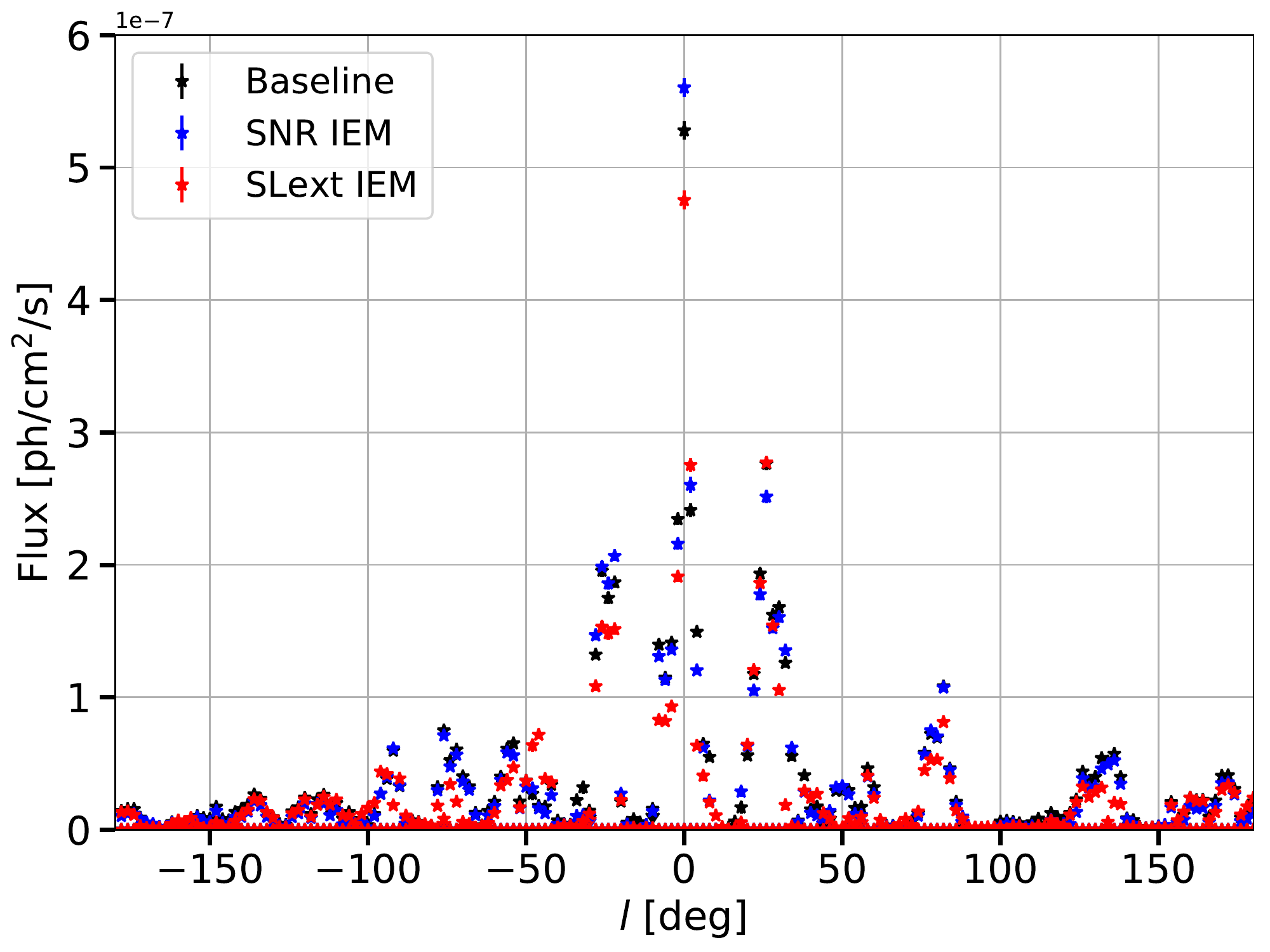}
\caption{Flux, integrated between $1-10$ GeV, absorbed by the DM template when it is placed at different positions along the Galactic plane. We show the results obtained with three different IEMs.}
\label{fig:GPsurvey}
\end{figure}

The presence of the GCE is robust over different data selection analysis techniques, IEMs and point source catalogs used to analyze the data (see, e.g., the results in this paper and \cite{Calore:2014xka,TheFermi-LAT:2017vmf}).
It is thus interesting to investigate if similar excesses are present in other directions towards the Galactic plane. 
In order to test this we use a DM template generated with $\gamma=1.2$
and we select different ROIs of size $40^{\circ}\times40^{\circ}$ and centered at different directions on the Galactic plane from $l=0^{\circ}$ to $360^{\circ}$ with a step size of $2^{\circ}$.
This gives 180 ROIs for which we apply the same analysis explained in Sec.~\ref{sec:analysis}.
After performing a fit to the ROI we save the flux of the DM template integrated between 1 and 10 GeV.
We show in Fig.~\ref{fig:GPsurvey} the flux absorbed by the DM template located at the different longitudes. We display the results obtained with the {\tt Baseline}, {\tt SLext} and {\tt SNR} IEM, but we find similar outcomes with the other models.
The next highest flux is detected at $l=\pm20^{\circ}$, with a decreased amplitude by a factor of $\sim 2$. 
These two excesses have been already found in \cite{Calore:2014xka} by performing a similar analysis, and they are probably associated to the $\gamma$-ray emission from molecular clouds. Our results are very similar to results found in Ref.~\cite{Calore:2014xka}.

\subsection{Energy spectrum}
\label{sec:spectrum}

\begin{figure}
\includegraphics[width=0.49\textwidth]{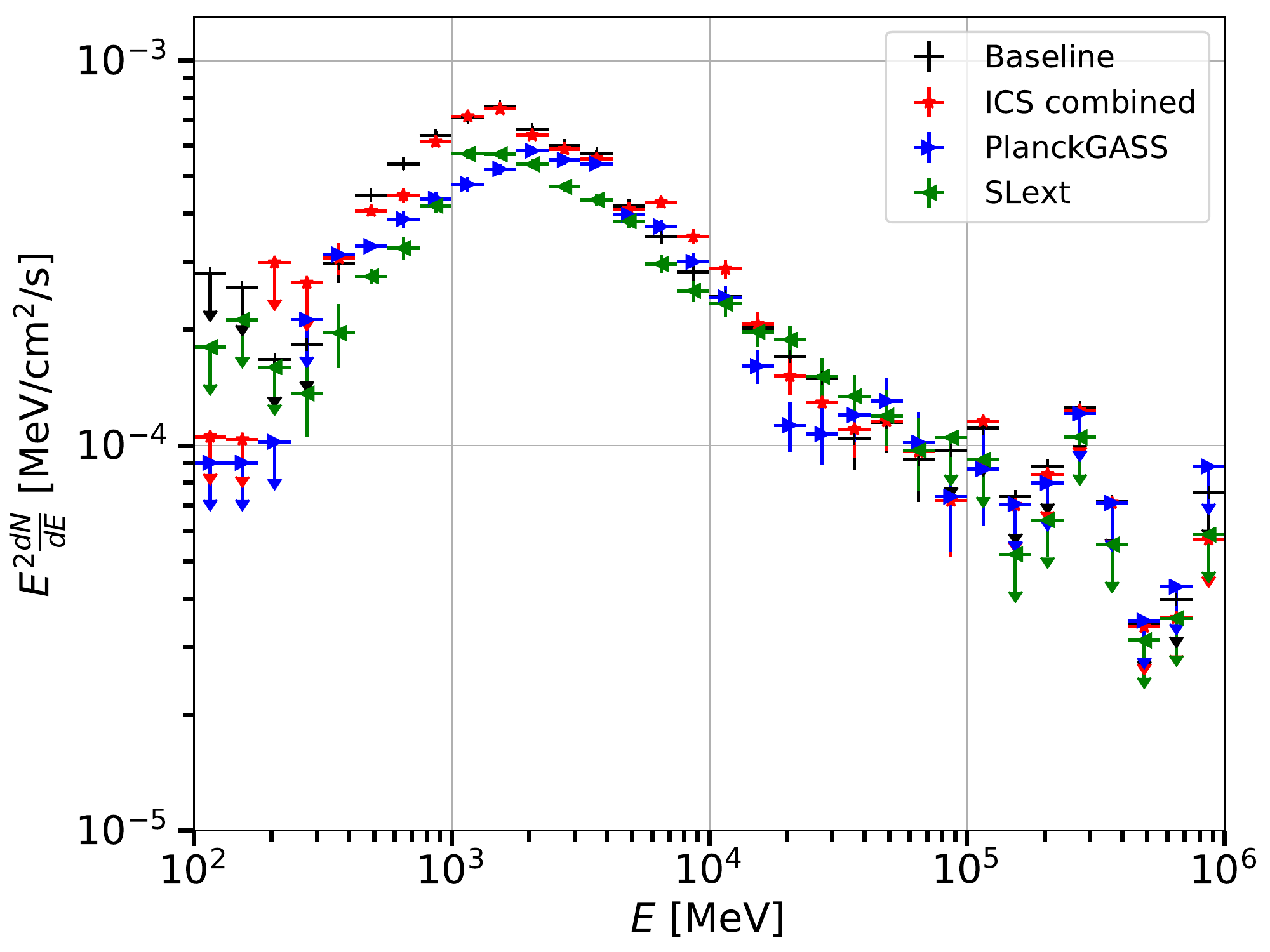}
\includegraphics[width=0.49\textwidth]{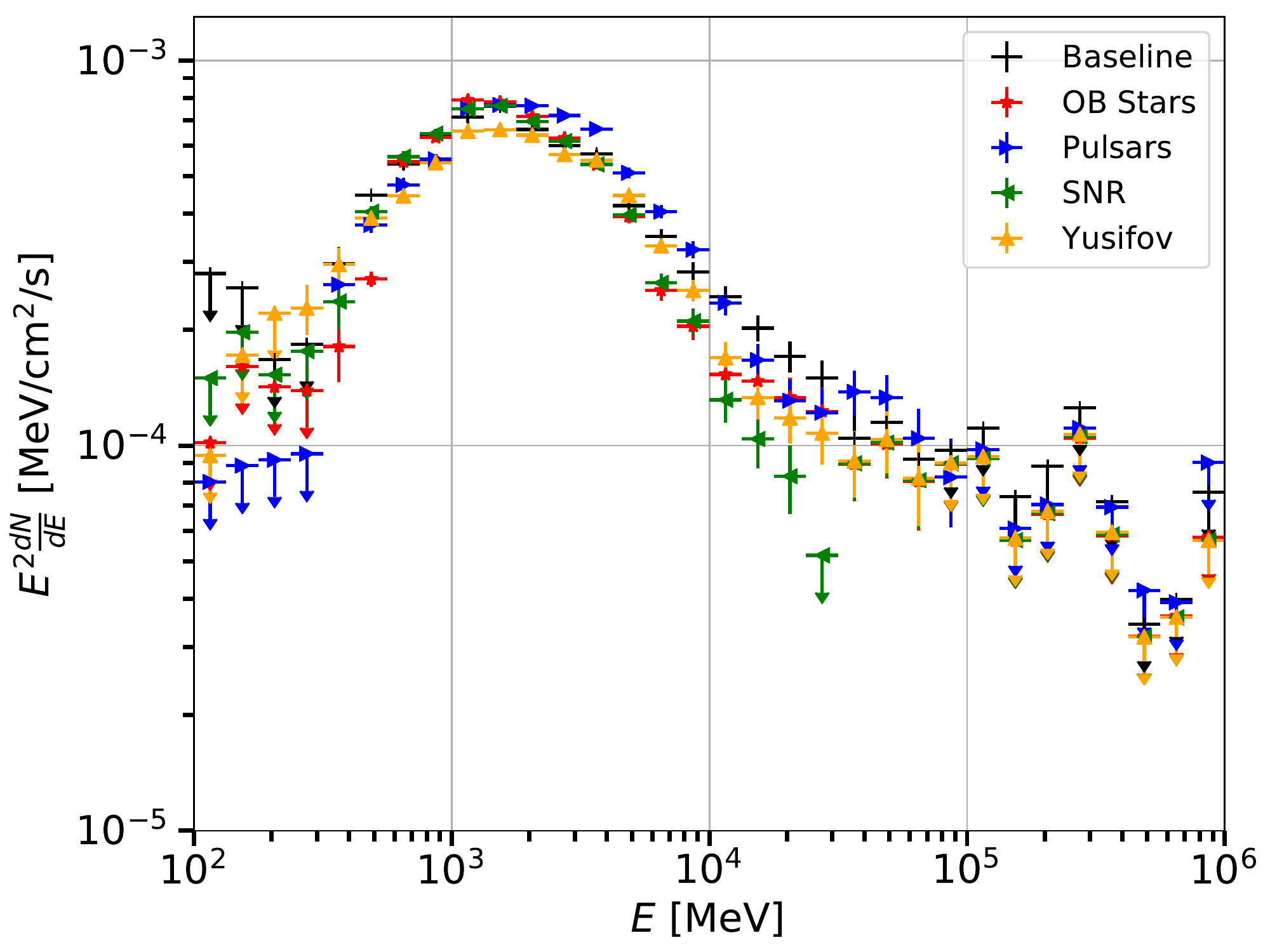}
\includegraphics[width=0.49\textwidth]{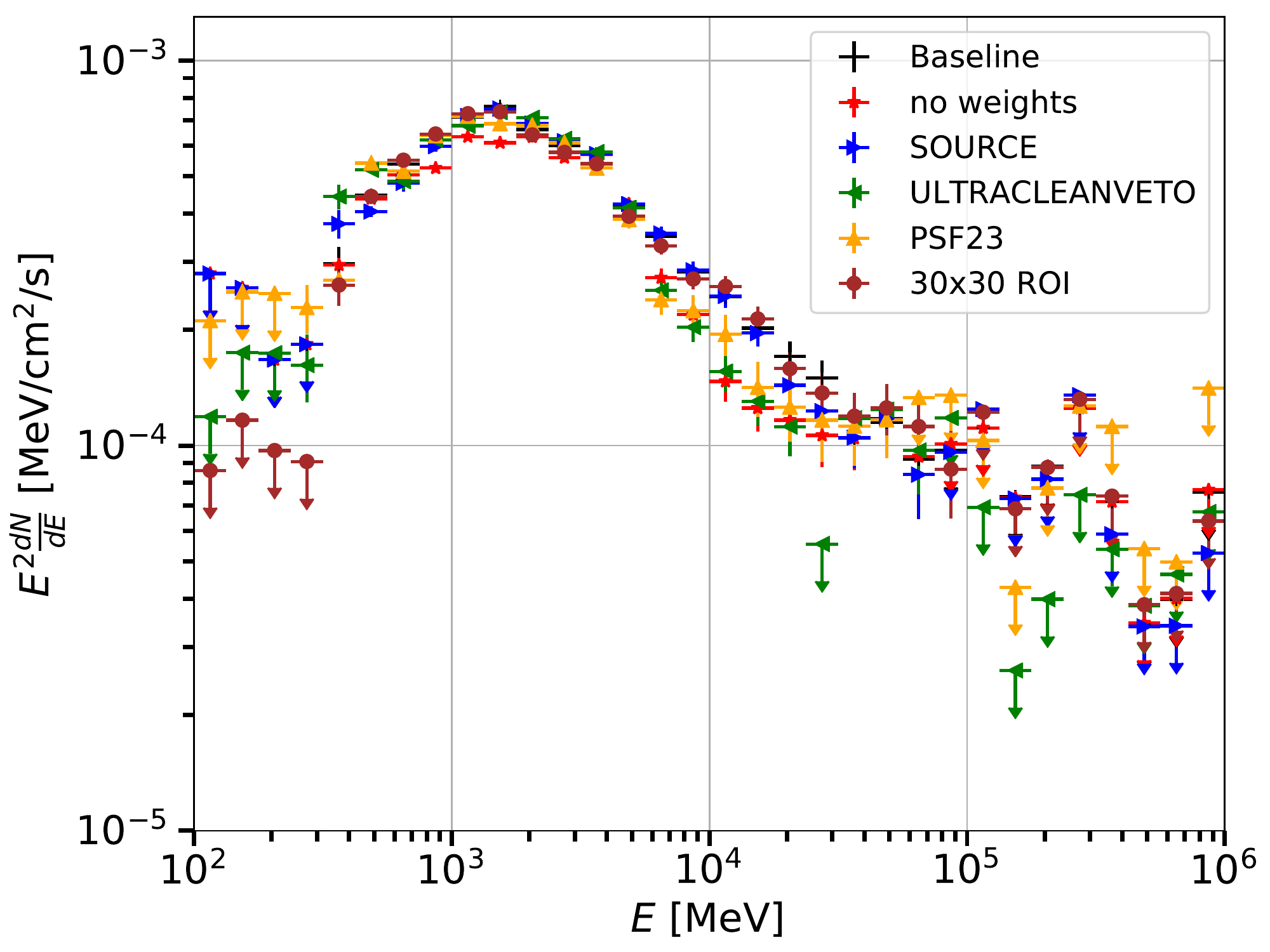}
\caption{GCE SED measured when using the {\tt Baseline}, {\tt Yusifov}, {\tt SNR}, {\tt OBstars}, {\tt Pulsars}, {\tt SLext}, {\tt ICS combined} and {\tt PlanckGASS} IEMs. We also show the results obtained with different selections of the data ({\tt SOURCE}, {\tt ULTRACLEANVETO} and {\tt PSF23}), with a smaller ROI ({\tt $30\times30$ ROI}) and without using the weight map in the maximum likelihood analysis ({\tt no weights}).}
\label{fig:spectrumcases}
\end{figure}

\begin{figure*}
\includegraphics[width=0.60\textwidth]{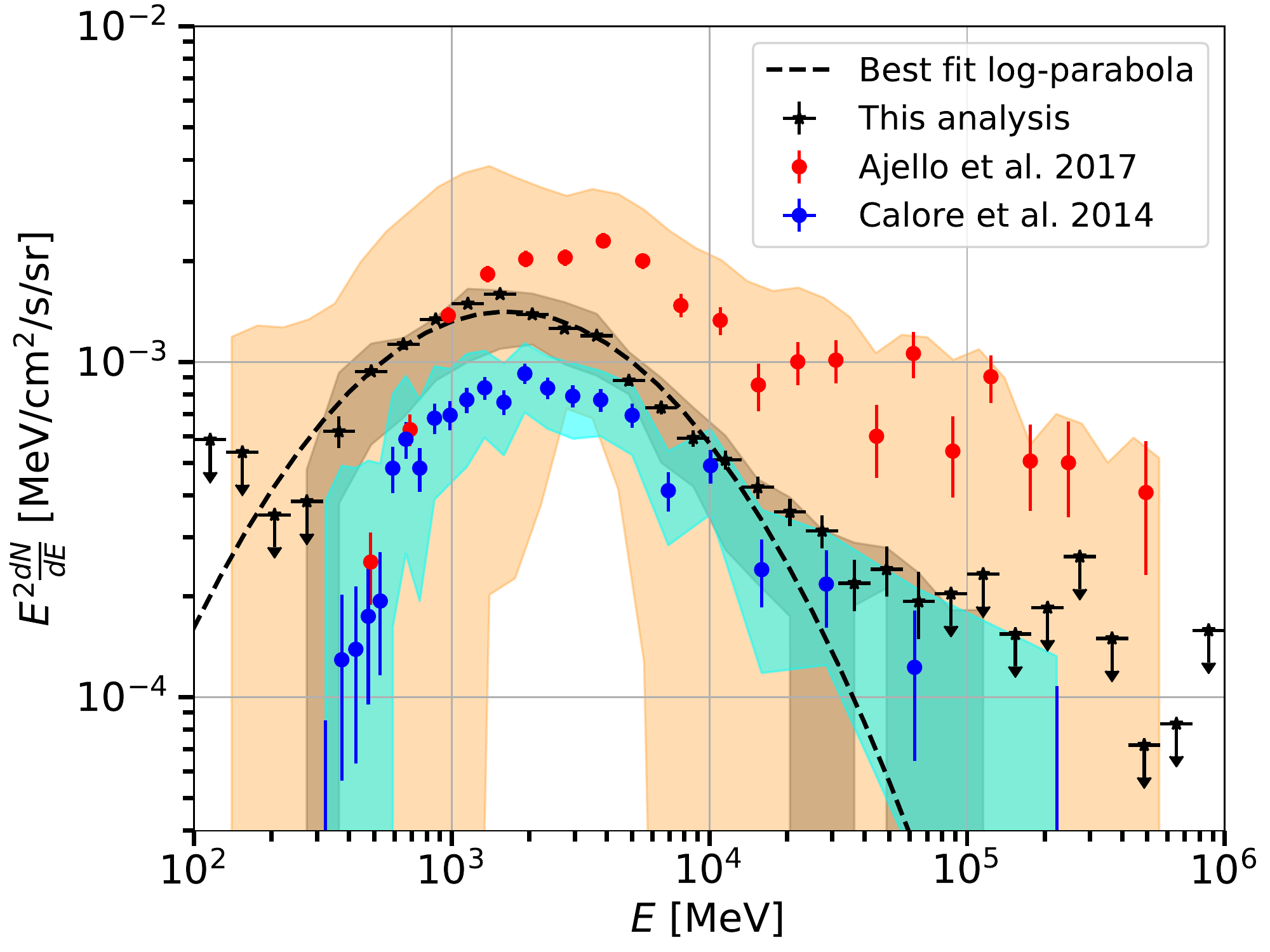}
\caption{Comparison between the results for the GCE SED obtained in our analysis and in \cite{Calore:2014xka,TheFermi-LAT:2017vmf}. The bands represent the variation of the GCE SED obtained by using all the IEMs and analysis techniques shown in Fig.~\ref{fig:spectrumcases} and the results found in \cite{Calore:2014xka,TheFermi-LAT:2017vmf} when using different IEMs. See the text for further details on the conversion of the GCE SED found in our analysis and in \cite{Calore:2014xka,TheFermi-LAT:2017vmf} into flux per solid angle (i.e., in units of MeV/cm$^2$/s/sr). We also display the best-fit to the GCE SED, obtained with the {\tt Baseline} IEM, by using a log-parabola function.}
\label{fig:spectrumconv}
\end{figure*}

\begin{figure}
\includegraphics[width=0.49\textwidth]{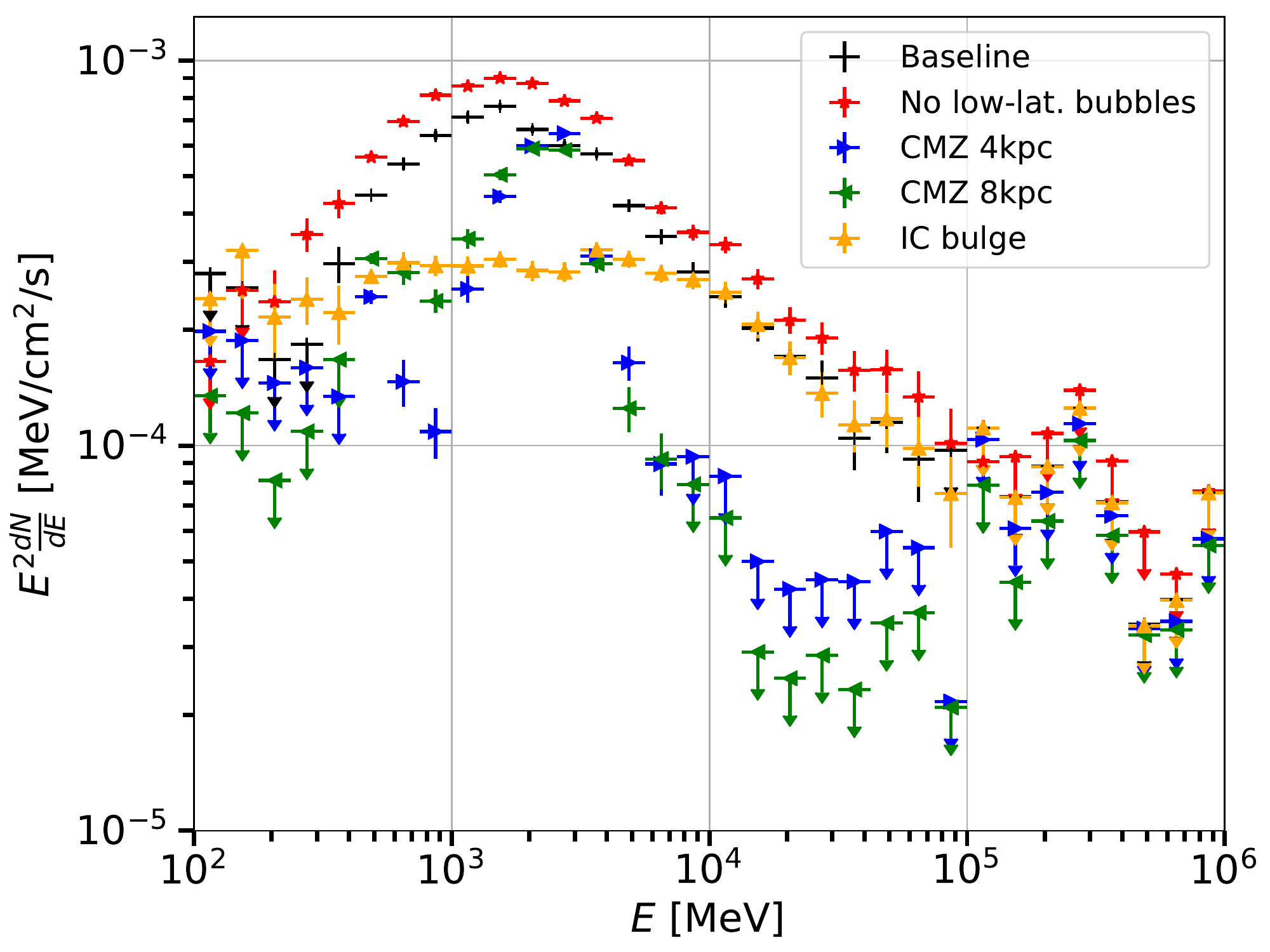}
\caption{GCE SED obtained in case we use the {\tt Baseline}, {\tt CMZ 4kpc}, {\tt CMZ 8kpc}, {\tt IC bulge} and {\tt no low-lat bubbles} IEMs.}
\label{fig:spectrumaddcomp}
\end{figure}

The first characteristic of the GCE that we investigate is the energy spectrum.
In order to calculate the GCE SED, we consider the energy range $0.1-1000$ GeV and we apply the analysis pipeline presented in Sec.~\ref{sec:analysistec}.

The GCE is detected with a maximum $TS$ of $1000-3000$, depending on the IEM, in the energy range between $0.8-3$ GeV.
These $TS$ values correspond to a significance of the GCE of about $40\sigma$, i.e., larger than that reported in \cite{Daylan:2014rsa}, where they find $30\sigma$ using the interstellar emission modeled with a unique template.
We find a larger significance for the GCE because we select a larger data sample (11 versus 5.5 years) and more refined models for the background components with respect to \cite{Daylan:2014rsa}.
At lower and higher energies the GCE $TS$ drops quickly, and for energies smaller than 0.3 GeV and larger than $30-50$ GeV the value goes below 25 for most of the IEMs.

We show in Fig.~\ref{fig:spectrumcases} the GCE SED found for the following IEMs: {\tt Baseline}, {\tt Yusifov}, {\tt SNR}, {\tt OBstars}, {\tt Pulsars}, {\tt SLext}, {\tt ICS combined} and {\tt PlanckGASS}. 
We also test different selections of the data ({\tt SOURCE}, {\tt ULTRACLEANVETO}, {\tt PSF23}), a smaller ROI ({\tt ROI $30\times30$}) and the exclusion of the weight maps in the maximum likelihood analysis ({\tt no weights}). 
All the results obtained with these IEMs, data selections, and techniques share the same general behavior. The GCE has a bumpy SED, with a peak at around 1-3 GeV, and a low and high-energy tail similarly to what found in Refs.~\cite{Abazajian:2012pn,Daylan:2014rsa,Calore:2014xka,TheFermi-LAT:2015kwa,TheFermi-LAT:2017vmf}.
Most of the data points below 500 MeV are upper limits. 
This is due both to the falling SED of the GCE below 1 GeV and to the weighted likelihood technique, which de-emphasizes the statistical weight of low-energy photons. This paper represents the first use of this technique to account properly for the systematic differences resulting from differing IEMs, which for this region and at these energies is by far the dominant component (see Fig.~\ref{fig:spectrumaddcomp}).
In Fig.~\ref{fig:spectrumconv} we show the envelope of the GCE SEDs for all of the above IEMs compared to the results found in \cite{Calore:2014xka,TheFermi-LAT:2017vmf}. 
The normalization of the GCE SED changes by roughly 60\% when using different IEMs, data selections and analysis techniques.
While these results are similar to those  obtained in Refs.~\cite{2015PhRvD..91l3010Z,Calore:2014xka}, they are more robust due to using twice as much data, a newer catalog with almost twice as many sources, and IEMs designed specifically for the Galactic center region.
The SED is shown in this figure in units of MeV/cm$^2$/s/sr, i.e., we have divided the SED by the ROI solid angle, since the selection of the ROI is different in our analysis with respect to \cite{Calore:2014xka,TheFermi-LAT:2017vmf}. In particular, Ref.~\cite{Calore:2014xka} selected a region given by  $2^{\circ}<|b|<20^{\circ}$ and $|l|<20^{\circ}$, while Ref.~\cite{TheFermi-LAT:2017vmf} used an ROI defined with an angular distance $<10^{\circ}$ from the Galactic center and a mask of bright sources (which implies an effective mask of $2^{\circ}$ from the Galactic center). Considering the systematic band due to the choice of the IEM, our results are compatible with the ones reported in \cite{Calore:2014xka,TheFermi-LAT:2017vmf}.

We fit different analytical functions to the GCE SED. In particular we test a power-law, a power-law with an exponential cutoff (PLE) and a log-parabola (LP)\footnote{\url{https://fermi.gsfc.nasa.gov/ssc/data/analysis/scitools/source_models.html#spectralModels}.}. 
We apply the fits to the SED measured with the {\tt Baseline} IEM. Similar results are found by fitting the SED obtained with the other models. 
We find a much better match with the data using the log-parabola shape and we find best-fit values for the spectral index of $-2.0$ and curvature index of 0.27. 
In particular the $TS$ calculated as $2(\rm{Log}{\mathcal{L}_{\rm{LP}}-\rm{Log}\mathcal{L}_{\rm{PLE}}})$, where $\rm{Log}{\mathcal{L}_{\rm{LP}}}$ ($\rm{Log}{\mathcal{L}_{\rm{PLE}}}$) is the fit obtained by using the LP (PLE) SED, is 380.
Ref.~\cite{Abazajian:2012pn} also found a preference for the fit with a LP with a roughly similar value for the spectral and curvature indexes.
We display in Fig.~\ref{fig:spectrumconv} the comparison between the data and the LP best-fit. 
A LP shape is able to reproduce the GCE spectrum between $0.1-10$ GeV but is not able to properly capture the high-energy tail.  However, this tail is not significant for all the IEMs (see Fig.~\ref{fig:spectrumconv} where the convolution of the results obtained for all the IEMs is displayed).

Possible interpretations of the GCE are associated with the $\gamma$-ray emission from cosmic-ray protons and/or electrons and positrons injected from the Galactic center.
We test these possibilities using the {\tt CMZ 4kpc}, {\tt CMZ 8kpc}, and {\tt IC bulge} models. 
We also run the analysis for the case without the presence of the low-latitude bubbles component ({\tt no low-lat bubbles}).
The results obtained with these models are presented in Fig.~\ref{fig:spectrumaddcomp}.
The case {\tt no low-lat bubbles}, as expected, provides a $20-30\%$ larger SED because the GCE absorbs part of the low-latitude bubble emission. This model fits much worse the ROI giving a $\rm{Log}{\mathcal{L}}$ lower by 2100 with respect to the {\tt Baseline} model.
In the cases {\tt CMZ 4kpc}, {\tt CMZ 8kpc}, {\tt IC bulge} we measure a smaller GCE flux since the additional cosmic-ray components take part, but not all, the GCE emission. 
The GCE SED changes significantly with these latter models, but an excess peaked at a few GeV still remains with a high significance.
In particular, the model for which the GCE spectrum decreases the most, roughly by a factor of two, is the {\tt IC bulge} case. 
This model represents the possible flux of a population of pulsars located around the bulge of our Galaxy. 
This result demonstrates the viability of the millisecond pulsar interpretation for the GCE \cite{Bartels:2015aea,Lee:2015fea}.

Since the cases {\tt CMZ 4kpc}, {\tt CMZ 8kpc}, {\tt IC bulge} absorb a significant fraction of the GCE and they have been considered in the past as possible interpretations to the GCE (see, e.g., \cite{Carlson:2014cwa,Petrovic:2014uda,Gaggero:2015nsa}), we have tested these model without including the DM template. Therefore, we try to fully explain the GCE with the $\gamma$-ray emission produced for inverse Compton scattering, bremsstrahlung or $\pi^0$ decays, by cosmic rays injected from the Galactic center during recent outbursts.
The fits provide differences of likelihood with respect to the {\tt Baseline} that are $+420$ and $+540$ for the models {\tt CMZ 4kpc}, {\tt CMZ 8kpc}, (the fit improves) and $-230$ for the model {\tt IC bulge} (the fit worsens). 
The models labeled as {\tt CMZ 4kpc} and {\tt CMZ 8kpc} consider two more components with respect to {\tt Baseline} because they substitute the DM component with three templates associated to the $\gamma$-ray emission for inverse Compton, Bremsstrahlung and $\pi^0$ decay. 
Therefore, it is difficult to compare the improved values of the $\rm{Log}(\mathcal{L})$ obtained with the {\tt CMZ 4kpc} and {\tt CMZ 8kpc} with respect to the {\tt Baseline} model that uses the DM template. 
Moreover, the additional components present in the {\tt CMZ 4kpc} and {\tt CMZ 8kpc}, associated with the Bremsstrahlung and $\pi^0$ decay emissions, follow the distribution of interstellar gas that is not spherically symmetric around the Galactic center but rather elongated on the Galactic plane. 
However, as we will see in the next sections, the GCE is spherically symmetric.
So we think the improvement in the $\rm{Log}(\mathcal{L})$ with these models is due to fitting better the emission from the Galactic plane rather than accounting better for the GCE emission.
We tested different scenario for the {\tt IC bulge} model by assuming different sizes for the diffusive halo with the vertical size $z$ varied between 4 to 10 kpc and the radius $R$ from 10 to 20 kpc. However, the result is very similar for all the tested cases: the {\tt IC bulge} model performs always worse than the {\tt Baseline} model with the DM template with a difference of $\rm{Log}(\mathcal{L})$ between -250 and -230.
The two main reasons for this result are the following. First, $\gamma$ rays produced from electrons and positrons injected from the Galactic bulge (this is the component that in the {\tt IC bulge} model absorbs the GCE) has a much flatter dependence with the angular distance from the Galactic center in the inner few degrees with respect to the GCE (see Sec.~\ref{sec:spacial} and Fig.~\ref{fig:SB}). Second, the inverse Compton emission from the Galactic bulge does not have a spherically symmetric morphology as we measure for the GCE (see Fig.~\ref{fig:fluxmap} and also the model used in \cite{Macias:2016nev}).
To conclude the {\tt IC bulge} model can fit reasonably well the GCE but with a worse overall fit compared to a DM profile generated with a NFW density with $\gamma\sim 1.25$.

\subsection{Spatial morphology}
\label{sec:spacial}

\begin{figure}
\includegraphics[width=0.49\textwidth]{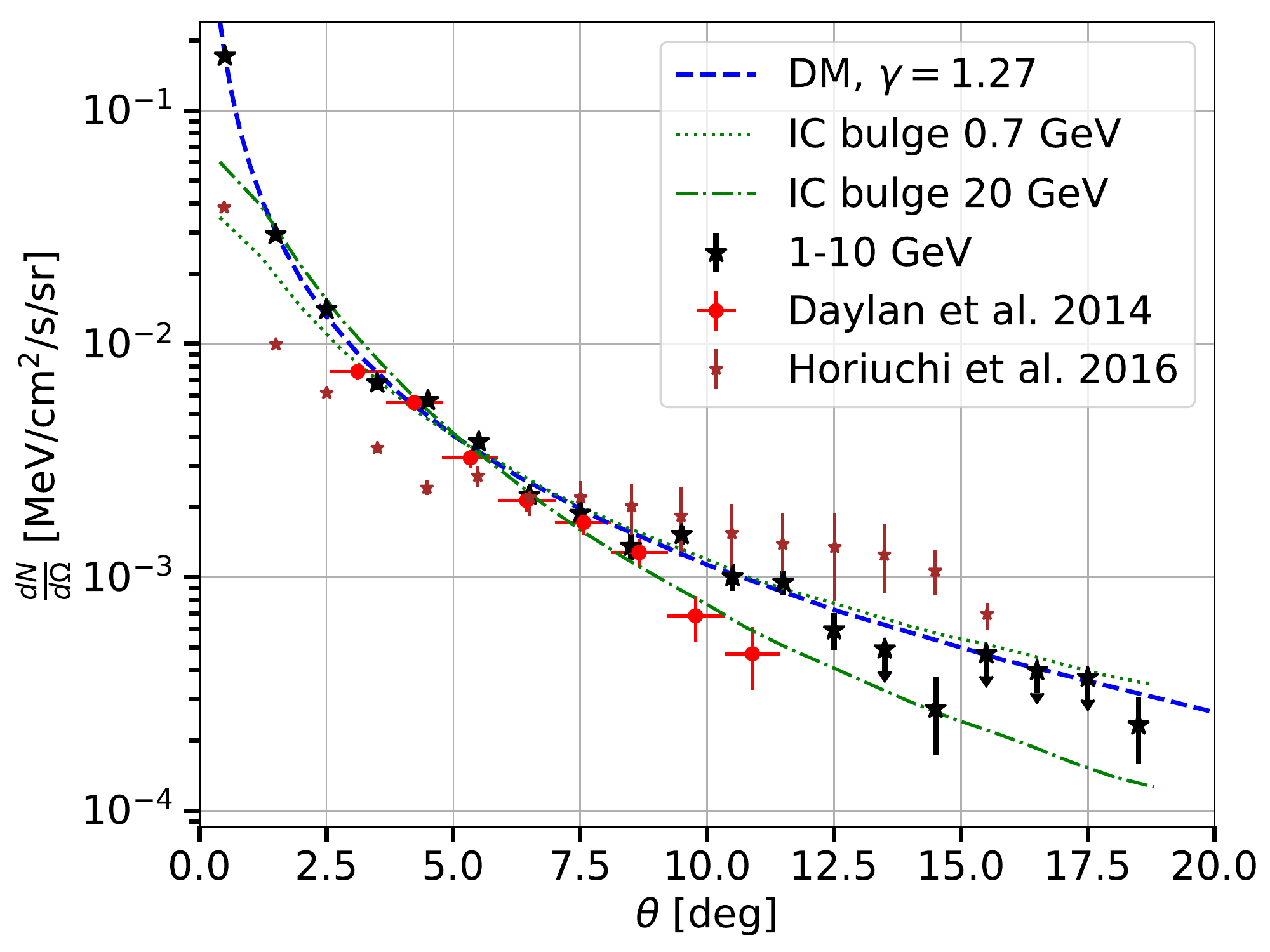}
\caption{Surface brightness data ($dN/d\Omega$) obtained with the analysis in annuli of size $1^{\circ}$ and with the {\tt Baseline} IEM in the energy range $1-10$ GeV. We also show the best fit obtained with a NFW density profile for $\gamma=1.27$ (dashed blue line) and for the {\tt IC bulge} model at $E=0.7$ (dotted green line) and 20 GeV (dot-dashed green line) for which we fit the GCE data with the $\gamma$ rays emitted for inverse Compton scattering by cosmic-ray electrons injected from the Galactic bulge (see Sec.~\ref{sec:analysisiem} for a complete description of this model). Red data points show the results obtained in \cite{Daylan:2014rsa} at 2.67 GeV.}
\label{fig:SB}
\end{figure}

\begin{figure}
\includegraphics[width=0.49\textwidth]{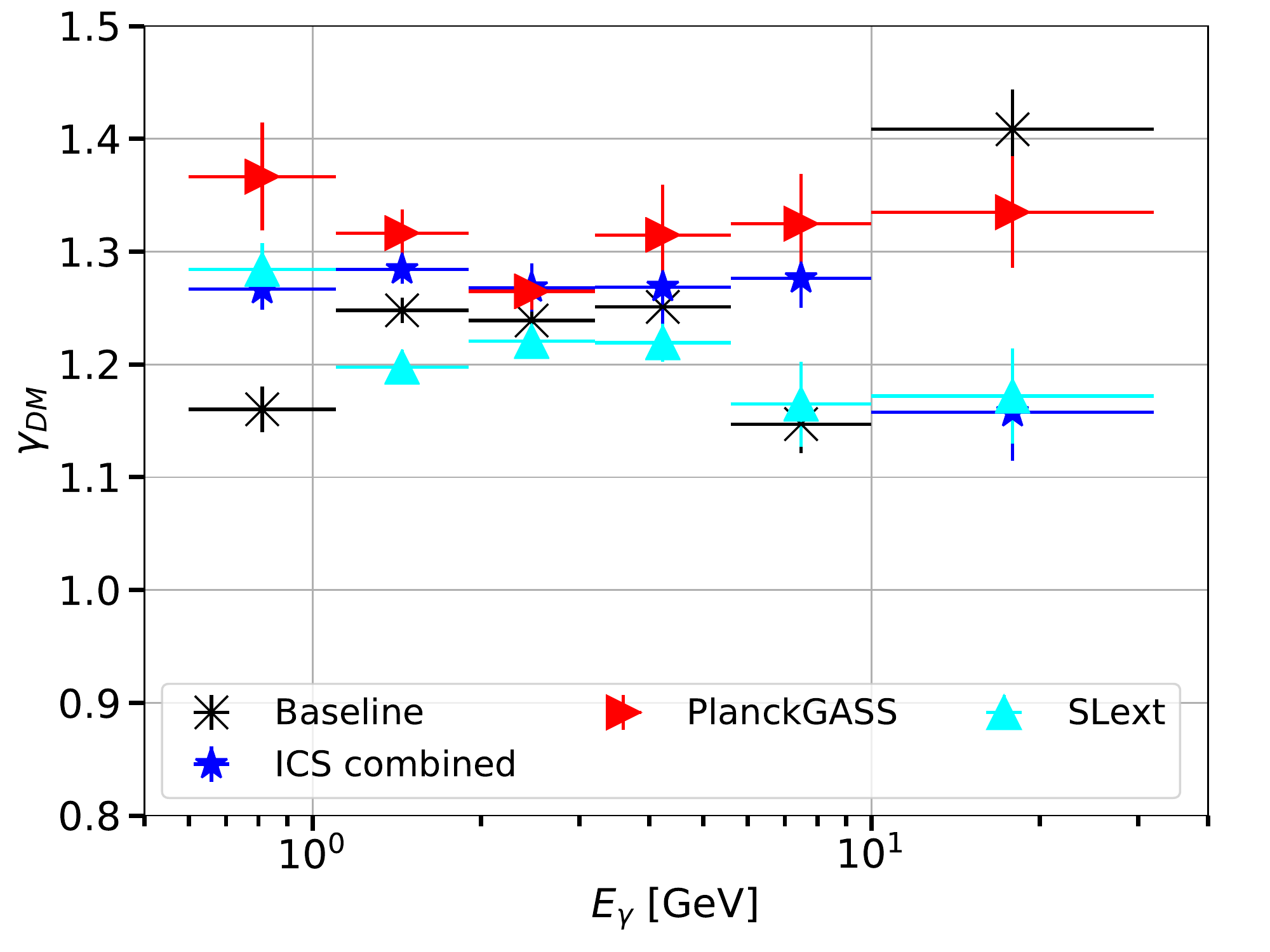}
\includegraphics[width=0.49\textwidth]{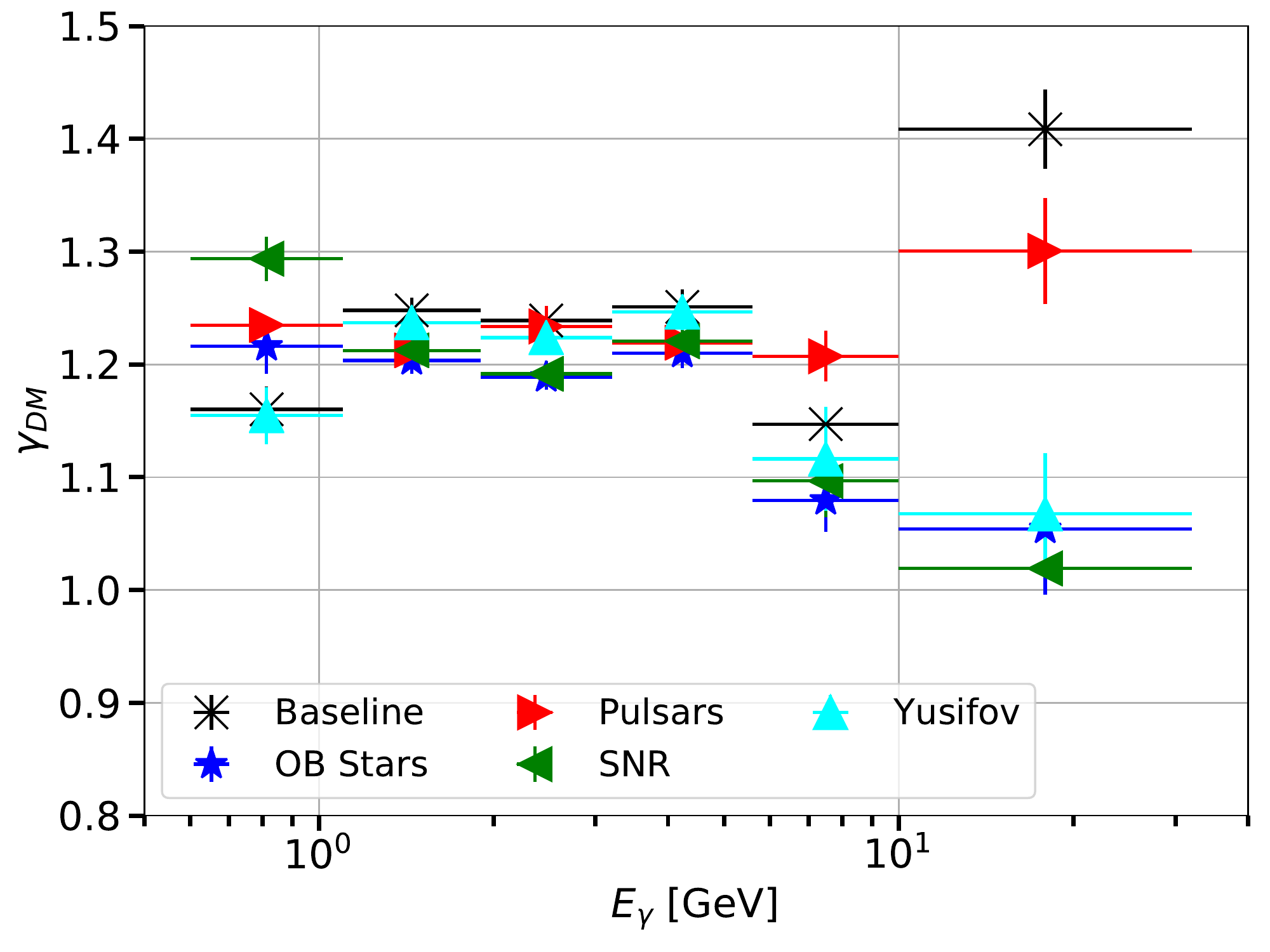}
\includegraphics[width=0.49\textwidth]{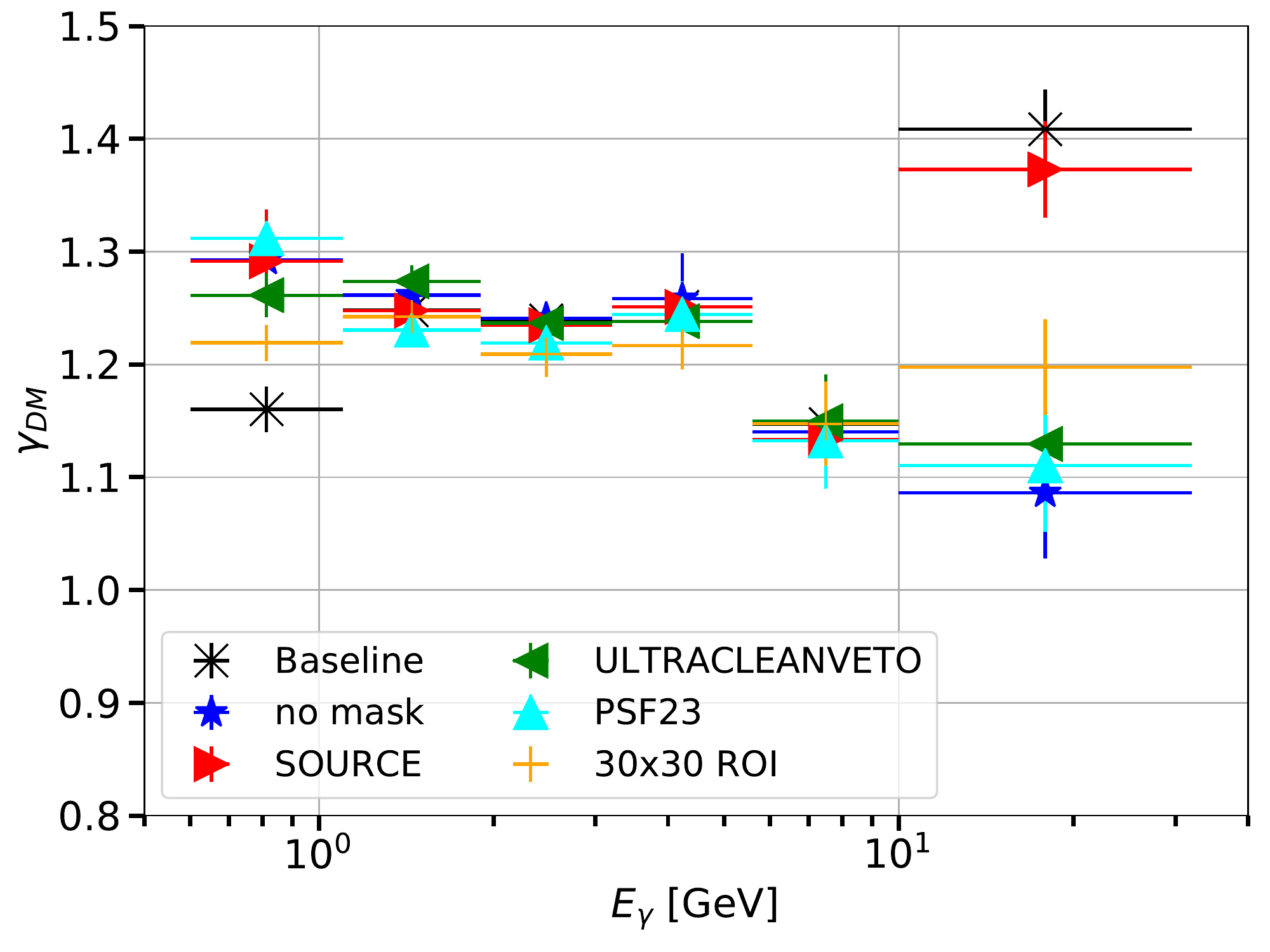}
\caption{Best fit for $\gamma$ obtained by fitting the surface brightness data obtained in the following energy bins: $0.6-1.0$, $1.0-1.8$, $1.8-3.2$, $3.2-5.6$, $5.6-10$, $10-32$ GeV. We show the results found using following IEMs: {\tt Baseline}, {\tt Yusifov}, {\tt SNR}, {\tt OBstars}, {\tt Pulsars}, {\tt SLext}, {\tt ICS combined} and {\tt PlanckGASS} IEM. We also display the values of $\gamma$ we obtained with different selections of the data ({\tt SOURCE}, {\tt ULTRACLEANVETO} and {\tt PSF23}), with a smaller ROI ({\tt $30\times30$ ROI}) and without using the weight map in the maximum likelihood analysis ({\tt no weights}).}
\label{fig:gammafit}
\end{figure}

In this paper we apply several methods to study the spatial morphology of the GCE.
In this section we employ a model dependent and a model independent technique.

In the model independent technique we substitute the DM template with concentric and uniform annuli.
Then, we fit the annuli to the data, using the pipeline presented in Sec.~\ref{sec:analysistec}, and we extract the energy flux of each annulus.
Finally, we divide the annulus energy fluxes for the solid angles and we obtain the surface brightness ($dN/d\Omega)$) of the GCE.
As demonstrated in \cite{Dimaurosim} with simulated data, the optimal annulus size is between $0.75^{\circ}-1.5^{\circ}$. 
Indeed, angular widths of these orders are similar to the angular resolution of the LAT for $E\sim1$ GeV\footnote{\url{https://www.slac.stanford.edu/exp/glast/groups/canda/lat_Performance.htm}} and they are small enough to capture the right spatial distribution of the GCE. 
We show here the results for an annulus size of $1^{\circ}$, but our conclusions do not change by using $0.75^{\circ}$ or $1.5^{\circ}$.

In Fig.~\ref{fig:SB} we show the surface brightness data obtained with an analysis in the energy range $1-10$ GeV using the {\tt Baseline} IEM.
The surface brightness data are very precise in the inner $10^{\circ}$ where the annuli are detected with at least $10\sigma$ significance and the precision of the data is between $2-10\%$.
The GCE extends with a significant flux roughly up to $12^{\circ}$. This demonstrates that our choice of an ROI with a size of $40^{\circ}\times40^{\circ}$ is appropriate.
We can fit well these data with a NFW DM profile with $\gamma=1.27$ (see Eq.~\ref{eq:NFW}).
Refs.~\cite{Abazajian:2012pn,Daylan:2014rsa,Calore:2014xka,TheFermi-LAT:2017vmf} found similar best-fit values for $\gamma$. However, most of those references only provide the value of $\gamma$ and not the data for the flux of the GCE as a function of angular distance from the Galactic center as we do.
The main new result of this paper is that we provide the spatial distribution of the GCE for a wide region and with a method that does not depend on the specific DM model. We report the surface brightness data that can be used to find which astrophysical interpretation is more suitable to explain the GCE spatial distribution. Refs.~\cite{Daylan:2014rsa,Horiuchi:2016zwu} are the only two publications that published results in a similar way as ours. However, Ref.~\cite{Daylan:2014rsa} provided the results for a limited region between $2.5^{\circ}-10^{\circ}$, with data that are not as precise as ours and tested only one IEM that was not designed for the Galactic center region. We compare our results with those obtained in \cite{Daylan:2014rsa} at 2.67 GeV in Fig.~\ref{fig:SB}. The surface brightness data are compatible between $2^{\circ}-7^{\circ}$, while smaller angular distances are not considered by Ref.~\cite{Daylan:2014rsa} and at larger angles their surface brightness deviates significantly both from our result and from the DM template predictions.
Instead, the result in Ref.~\cite{Horiuchi:2016zwu} have been derived using almost one half of the data and the older Pass 7 data selection. The authors have considered three IEMs taken from \cite{Calore:2014xka}. The surface brightness they obtain for the energy range $E\in[1.9,10]$ GeV is shown in Fig.~\ref{fig:SB}. We have rescaled their measurements to the energy range 1-10 GeV to be comparable with ours\footnote{We have calculated the LogParabola best fit to the GCE SED (see Fig.~\ref{fig:spectrumcases}) and integrated it between $E\in[1.9,10]$ GeV and $E\in[1.0,10]$ GeV. Then, we taken the ratio between the two integrals to rescale the results in Ref.~\cite{Horiuchi:2016zwu}, that are given for $E\in[1.9,10]$ GeV, to our energy range.}. The GCE spatial distribution found in Ref.~\cite{Horiuchi:2016zwu} is significantly different from ours in the inner few degrees from the Galactic center. In fact it is compatible with a NFW profile with $\gamma\sim 1.0$, i.e.~smaller than most of the $\gamma$ values found in several other references.

We also test the model {\tt IC bulge} for which we eliminate the DM template and we fit the surface brightness data with $\gamma$ rays emitted for inverse Compton scattering by cosmic-ray electrons injected from the Galactic bulge (see Sec.~\ref{sec:analysisiem} for a complete description of this model). We perform the fit using the model at $E=0.7$ and 20 GeV to demonstrate that this emission mechanism would produce a signal with a spatial morphology that evolves with energy. In particular at 0.7 GeV the signal is flatter and it decreases by a factor of about 30 between $0^{\circ}$ and $10^{\circ}$ instead at 20 GeV changes by roughly 100 times. 
As we will see next in this section, the spatial morphology of the GCE does not change significantly between 0.6 and 30 GeV so the variation of the Galactic bulge emission as a function of energy is problematic for the interpretation of the GCE with this mechanism. 

We now explore a novel analysis, never made before, and derive the surface brightness data in different energy bins between $0.1-1000$ GeV to investigate if there is an energy dependence of the spatial morphology of the GCE. We do so by performing the same analysis presented before in the following energy bins : $0.1-0.3$, $0.3-0.6$, $0.6-1.0$, $1.0-1.8$, $1.8-3.2$, $3.2-5.6$, $5.6-10$, $10-32$, $32-100$, $100-1000$ GeV.
Exactly as for the energy range $1-10$ GeV, we find a good fit to the data with a DM profile for all the energies between $0.6-30$ GeV for which we detect significantly the GCE.
Instead, at energies lower than 0.6 GeV and higher than 30 GeV the GCE is not significant and so the analysis provides surface brightness data with upper limits for most of the annuli.
Therefore, the fit to the $dN/d\Omega$ at these energies is prohibitive.

For the analyses at energies $0.6-30$ GeV, we decide to study the possible energy dependence of the spatial morphology by fitting with a DM template the surface brightness data found into different energy bins. Indeed, we find that the surface brightness data are well fit with an NFW profile with $\gamma$ free to vary (similarly to the fit depicted in Fig.~\ref{fig:SB} for energies between $1-10$ GeV).
In Fig.~\ref{fig:gammafit} we report the best-fit $\gamma$ values obtained for different IEMs and selections of the data and ROI size.
We do not find an energy evolution of $\gamma$ and its value at energies between $1-10$ GeV ranges between $1.1-1.2$.
This scatter is larger than the error on the single measurements which is of about a few \%. 
This implies that the systematics on the choice of the IEM or data selection and analysis technique dominates the uncertainty on the value of $\gamma$.
On the other hand, the scatter in the average values obtained for each case considered is between $1.1-1.3$.
The global average value calculated considering all the IEM and analysis setup is 1.25. 

We can interpret the differences for the $\gamma$ values we obtain for the different cases as the systematic uncertainty for the GCE spatial distribution.
In \cite{Dimaurosim} we have studied this extensively with simulations, finding that the systematic uncertainties on $\gamma$ are on the order of $5\%$ from the injected value.
Using the real data we find roughly $10\%$ (obtained as $100\cdot 0.5\cdot(1.08-1.32)/1.25$), thus the systematics are twice larger than the one obtained in \cite{Dimaurosim}.
This difference is probably due to the fact that it is difficult to bracket with simulations all the possible systematic effects present in the real data.

We also use a model dependent technique to find the spatial morphology of the GCE that consists of fitting the GCE with DM templates generated with different values of $\gamma$. For each run with a different $\gamma$ value the background components (i.e., sources and IEM components) are the same so we are able to produce a profile of $\rm{Log}(\mathcal{L})$ as a function of $\gamma$.
By maximizing the function $\rm{Log}(\mathcal{L})(\gamma)$ we then find best-fit value for $\gamma$.
We apply this analysis to all the cases presented in Fig.~\ref{fig:gammafit}. The values of $\gamma$ are consistent with the ones derived with the fit to the surface brightness data (see Fig.~\ref{fig:gammafit}).

To conclude, we do not find any significant evolution in the value of $\gamma$, i.e., in the GCE spatial morphology. 
There is a scatter of the average values between $1.2-1.3$ that is due to different results obtained for the models and analysis setup considered in the analysis. 
A change in the $\gamma$ value due to an evolution of the GCE spatial morphology is not excluded for some models, as for the {\tt Baseline} for which $\gamma$ increases from 0.6 GeV to 30 GeV from 1.15 to 1.4. However, this evolution is not present for the other models tested. 
Considering all the cases reported in Fig.~\ref{fig:gammafit}, the change of $\gamma$ is constrained to be at most $\sim 10\%$ from the average value. 

The scatter on the value of $\gamma$ and a maximum evolution of $10\%$ in its value is smaller than the evolution expected if the GCE is due to electrons and positrons injected from the Galactic bulge (this component is included in the {\tt IC bulge} model). As we have demonstrated in \cite{Dimaurosim}, with this mechanism the GCE spatial morphology would significantly evolve with energy and $\gamma$ would increase by about a factor of $30\%$ from 0.6 to 30 GeV. 
Moreover, as we have reported in Fig.~\ref{fig:SB}, it is challenging with an electron and positron burst signal to produce a surface brightness compatible with the GCE data in the inner few degrees from the Galactic center.

Our results for the value of $\gamma$ are compatible with the ones published in \cite{Calore:2014xka,Daylan:2014rsa}, while in Ref.~\cite{TheFermi-LAT:2017vmf} they find slightly smaller values of the order of $1.0-1.1$.

\subsection{Analysis in quadrants}
\label{sec:quadrants}

\begin{figure}
\includegraphics[width=0.49\textwidth]{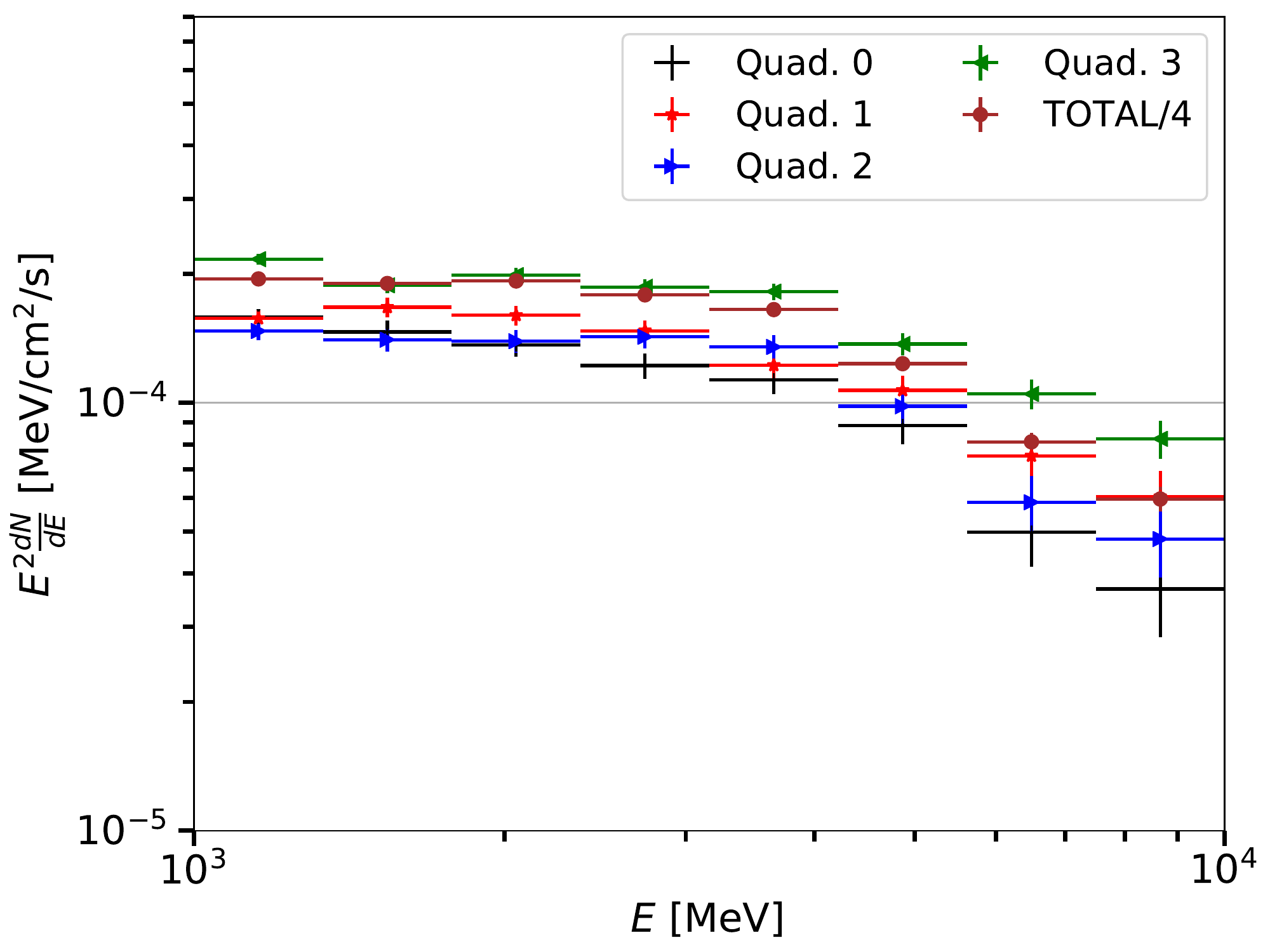}
\includegraphics[width=0.49\textwidth]{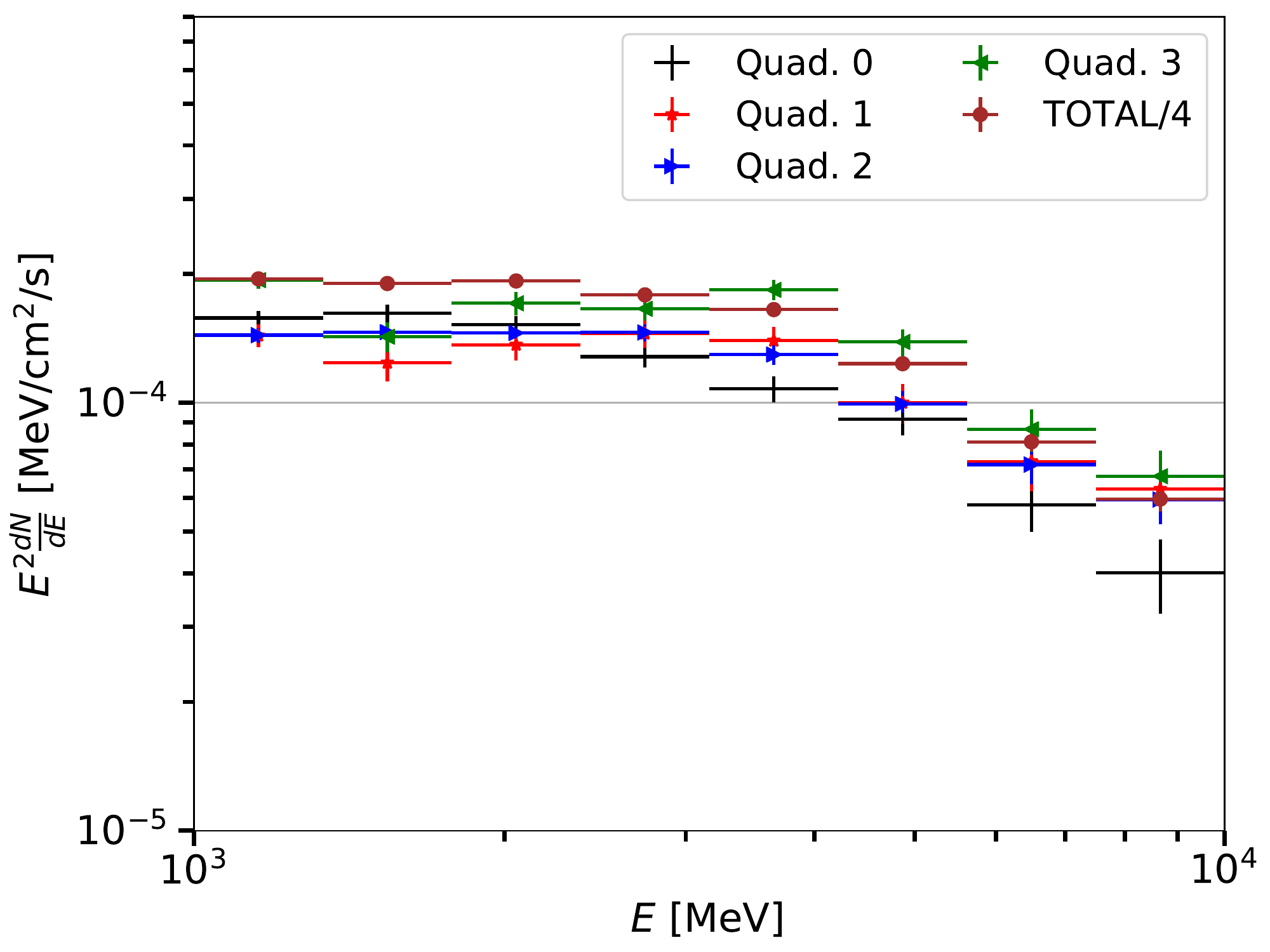}
\caption{SED of the quadrants for the {\tt quad$+$} configuration in the top panel and for {\tt quadx} in the bottom panel. We also display the GCE SED, obtained with the DM template considered as a whole, divided by 4. The data have been obtained with the {\tt Baseline} IEM.}
\label{fig:quadrantsSED}
\end{figure}

\begin{figure}
\includegraphics[width=0.49\textwidth]{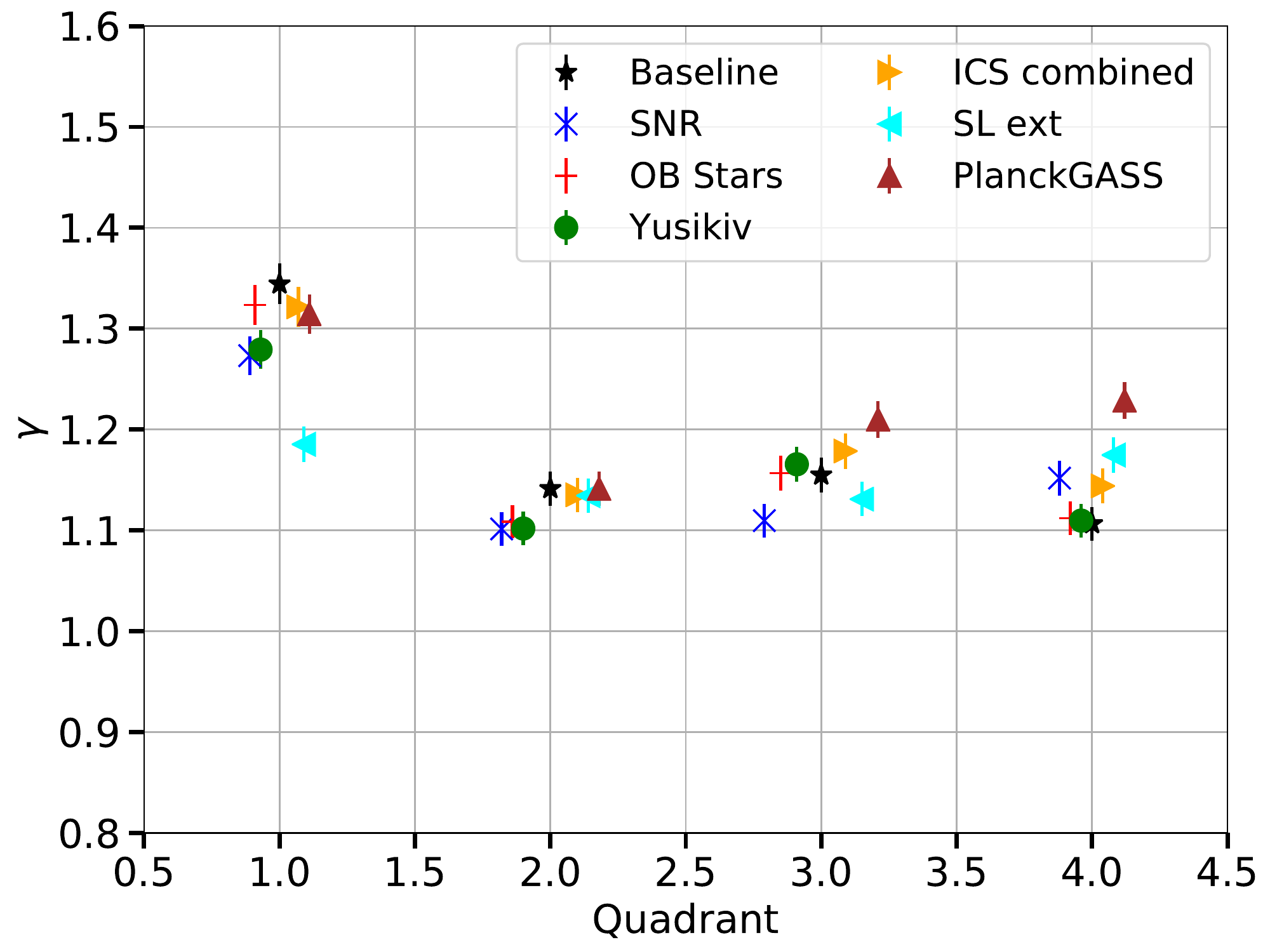}
\caption{Best-fit for $\gamma$ found by fitting the surface brightness of each quadrant and using different IEMs. We show here the results obtained with the quadrant configuration {\tt quad$+$}.}
\label{fig:quadrantsSB}
\end{figure}

We apply another method to study the spatial morphology of the GCE.
We divide the DM template into quadrants and we leave free to vary in the fit their SED parameters.
We focus this analysis in the energy range between $1-10$ GeV since we have not found any energy dependence of the value of $\gamma$, and in this energy range the GCE is detected with the largest significance.
We use a LP SED for each quadrant, thus the GCE is fit in this case with 12 SED parameters.
We test two different orientations of the quadrants.
The first, labeled as {\tt quad$+$}, has quadrants separated by the Galactic plane direction and by $b=\pm90^{\circ}$. The first quadrant is located at positive longitudes and latitudes, and the other quadrants move from the first in a counterclockwise direction. 
The second configuration, labeled as {\tt quadx}  is rotated by $+45^{\circ}$ with respect to the former, with the first quadrant located at positive longitudes and latitudes in the range $-45^{\circ}<b<45^{\circ}$. The second, third and fourth quadrants are found by moving from the first in a counterclockwise direction. Therefore, in the  {\tt quad$+$} configuration all the quadrants share a portion of the Galactic plane while in the {\tt quadx}, only the first and third quadrants include the Galactic plane. 
We use both these quadrant configurations to investigate a possible change in the results due to the choice of the IEM.

In Fig.~\ref{fig:quadrantsSED} we show the SED we find for each quadrant compared to the SED obtained when we use a unique DM template for the GCE and divide the flux value by four.
We report the result for the {\tt Baseline} IEM, but similar conclusions are found when taking the other models and different data selections.
All the quadrants share a similar SED and the difference in normalization is roughly $30\%$. 
The data are well fit with a LP with values for the spectral index $\alpha$ and curvature index $\beta$ compatible with the values found for the GCE when fitting it with a unique DM template (i.e., $\alpha=2.0$ and $\beta=0.27$).
There is not a relevant difference in the results considering the {\tt quad$+$} or {\tt quadx} configuration.
Finally, the SED of DM considered as a whole and divided by four is compatible with the quadrant SED considering the scatter.
In Ref.~\cite{Dimaurosim} we have performed a similar analysis with simulated data finding that the scatter is roughly $15\%$.
Therefore, also in this case the scatter that we find in the real data is larger, roughly by a factor of 2, compared to simulations .  
In Ref.~\cite{TheFermi-LAT:2017vmf} the authors performed a similar analysis finding that, when accounting for the asymmetric emission from the {\it Fermi} bubbles, the SED of the four quadrants are similar to each other. Our result is thus similar to theirs. 
Ref.~\cite{Calore:2014xka} inspected the GCE SED by considering ten different portions of the region around the Galactic center and they found similar spectra in all regions.
Therefore, we confirm the results found by previous publications: the GCE has the same spectrum also when it is divided into multiple spatial components.

We also study the spatial morphology of the quadrants by substituting the DM template with uniform annuli divided into quadrants.
Then, by fitting all the annuli quadrants to the data we find the surface brightness data of each quadrant.
Finally, we fit the surface brightness data of each quadrant with a DM template with $\gamma$ free to vary.
We perform this analysis using the {\tt quad$+$} quadrant configuration.
In Fig.~\ref{fig:quadrantsSB} we show the best-fit $\gamma$ we find for each quadrant if we run the analysis with different IEMs.
The results found are similar among the IEMs employed.
We find $\gamma\sim 1.15$ for the quadrants 2,3,4 and $\gamma\sim 1.3$ for the first quadrant. 
The only IEM that gives practically the same $\gamma$ for all quadrants is the {\tt SL ext}.
The differences we see in Fig.~\ref{fig:quadrantsSB} are due to larger negative residuals present in the first quadrant with respect to the others, that gives also a different spatial morphology for the DM template in this region of the Galactic center.
However, the difference in $\gamma$ is of the same order as that found when performing the analysis in annuli with different IEMs (see Fig.~\ref{fig:gammafit}).
Moreover, DM profiles with $\gamma$ of 1.15 and 1.3, if we normalize both cases to have the same flux at $1^{\circ}$, produce a difference at an angular distance of $2^{\circ}$ ($4^{\circ}$) from the Galactic center of roughly 20\% (40\%) between the two cases.
Thus these results do not indicate any significant asymmetry in the GCE. 
We will make a further test of the GCE symmetry in Sec.~\ref{sec:position}.

\subsection{Position}
\label{sec:position}

\begin{figure}
\includegraphics[width=0.49\textwidth]{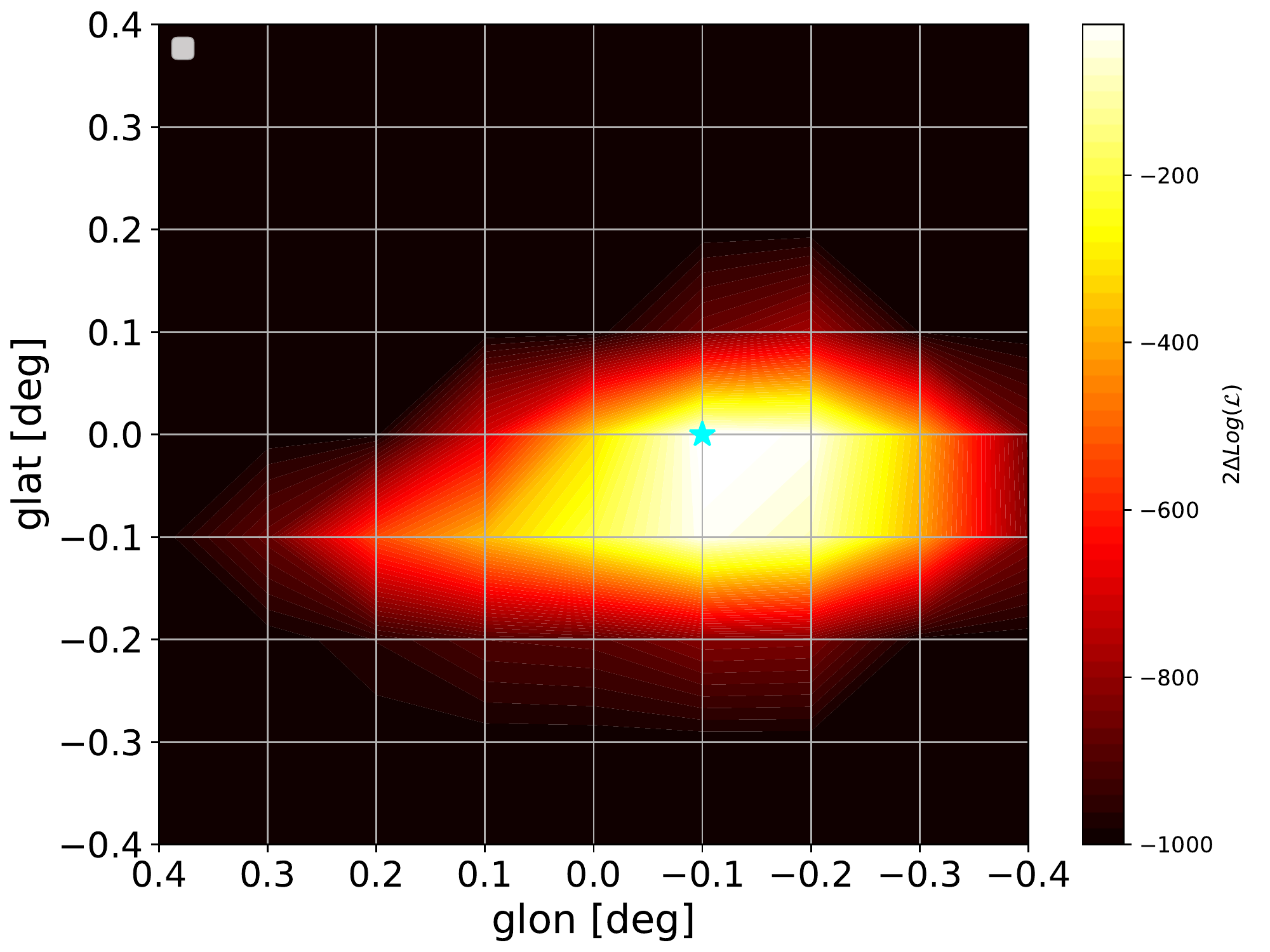}
\caption{Value of the likelihood as a function of the Galactic longitude and latitude $\rm{Log}(\mathcal{L})(l,b)$ found with the {\tt Baseline} IEM for the energy range between $1-10$ GeV. We show with a cyan star the position for which we maximize the $\rm{Log}(\mathcal{L})(l,b)$.}
\label{fig:pos}
\end{figure}

\begin{figure*}
\includegraphics[width=0.49\textwidth]{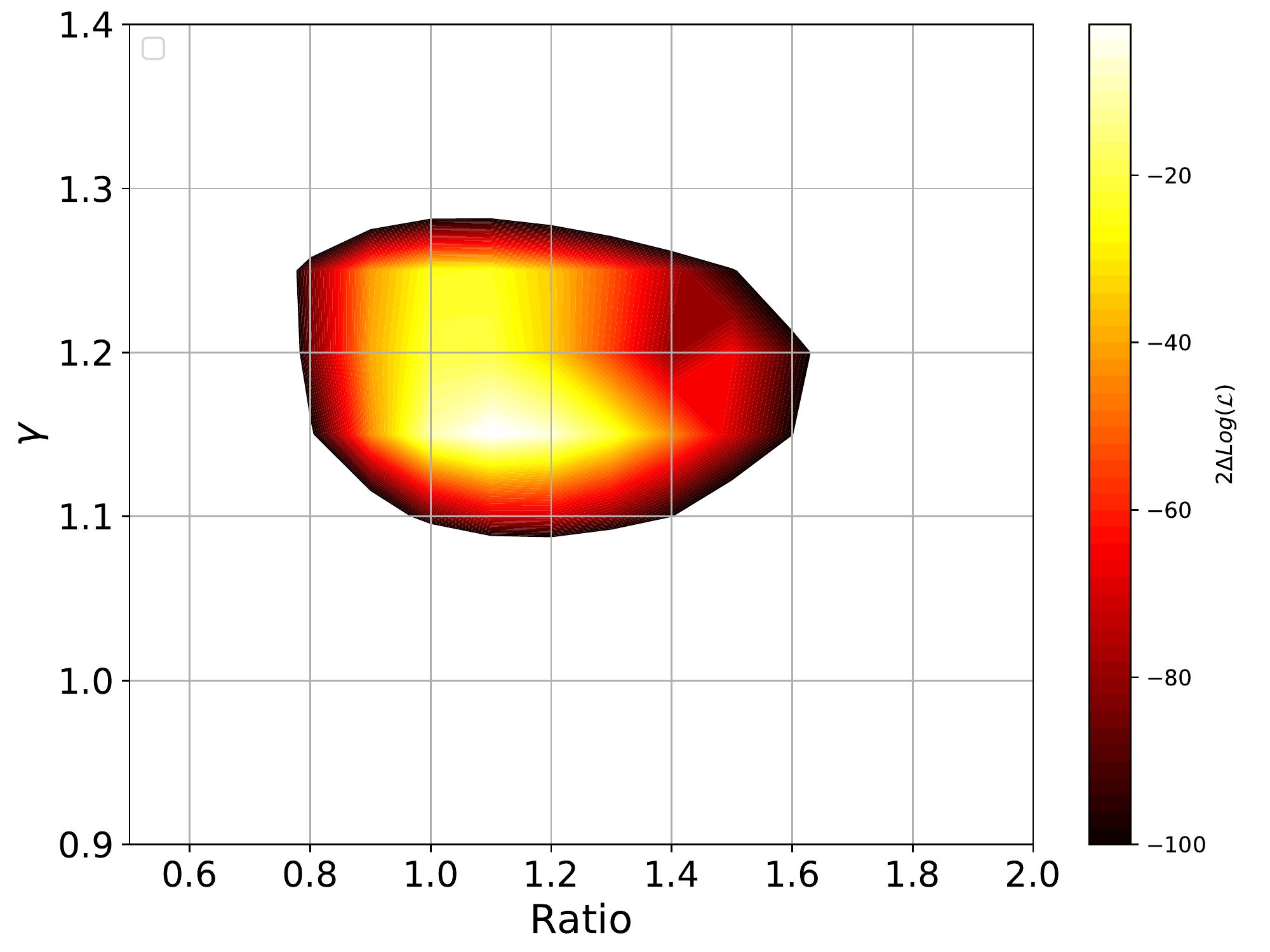}
\includegraphics[width=0.49\textwidth]{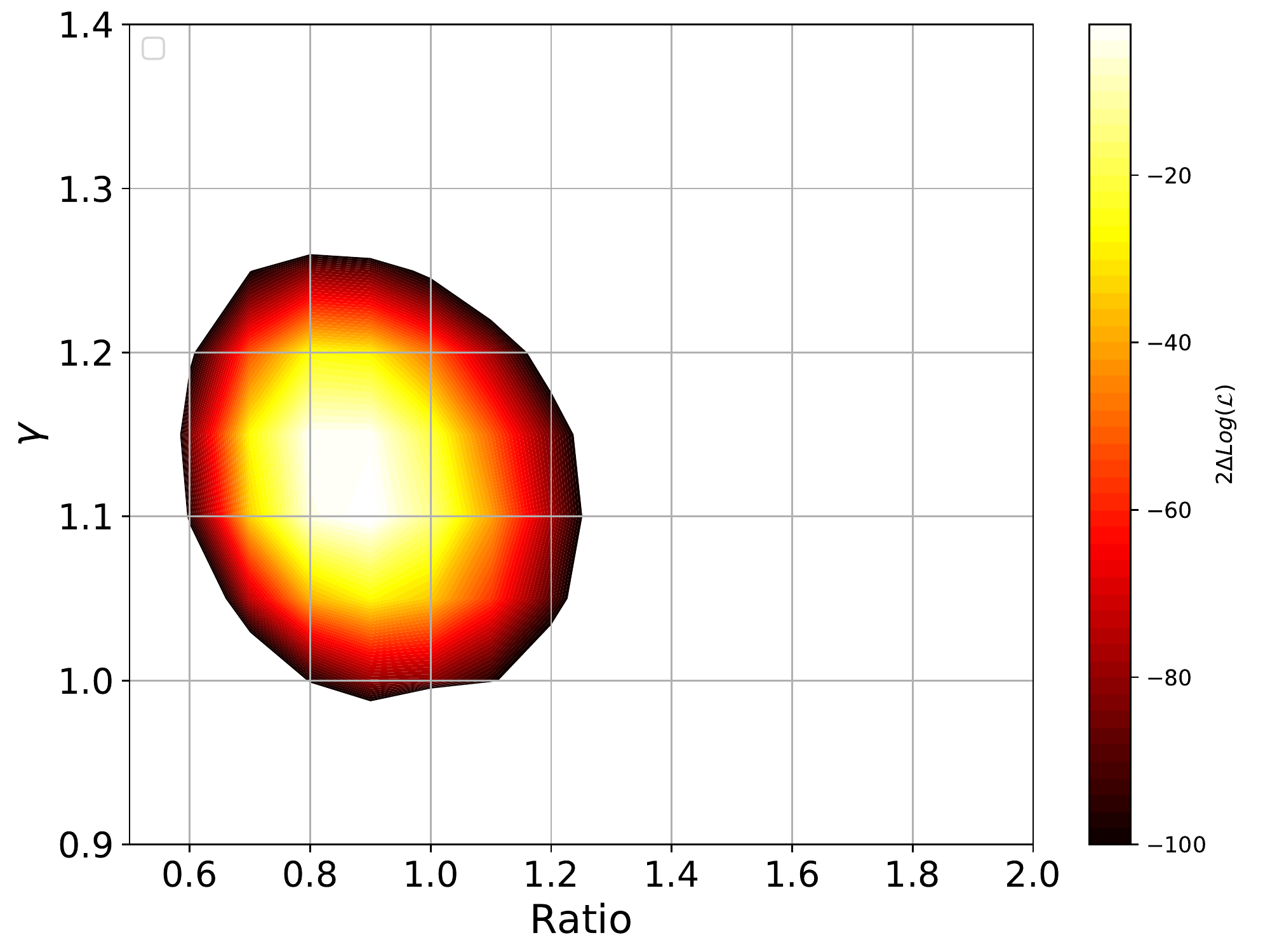}
\includegraphics[width=0.49\textwidth]{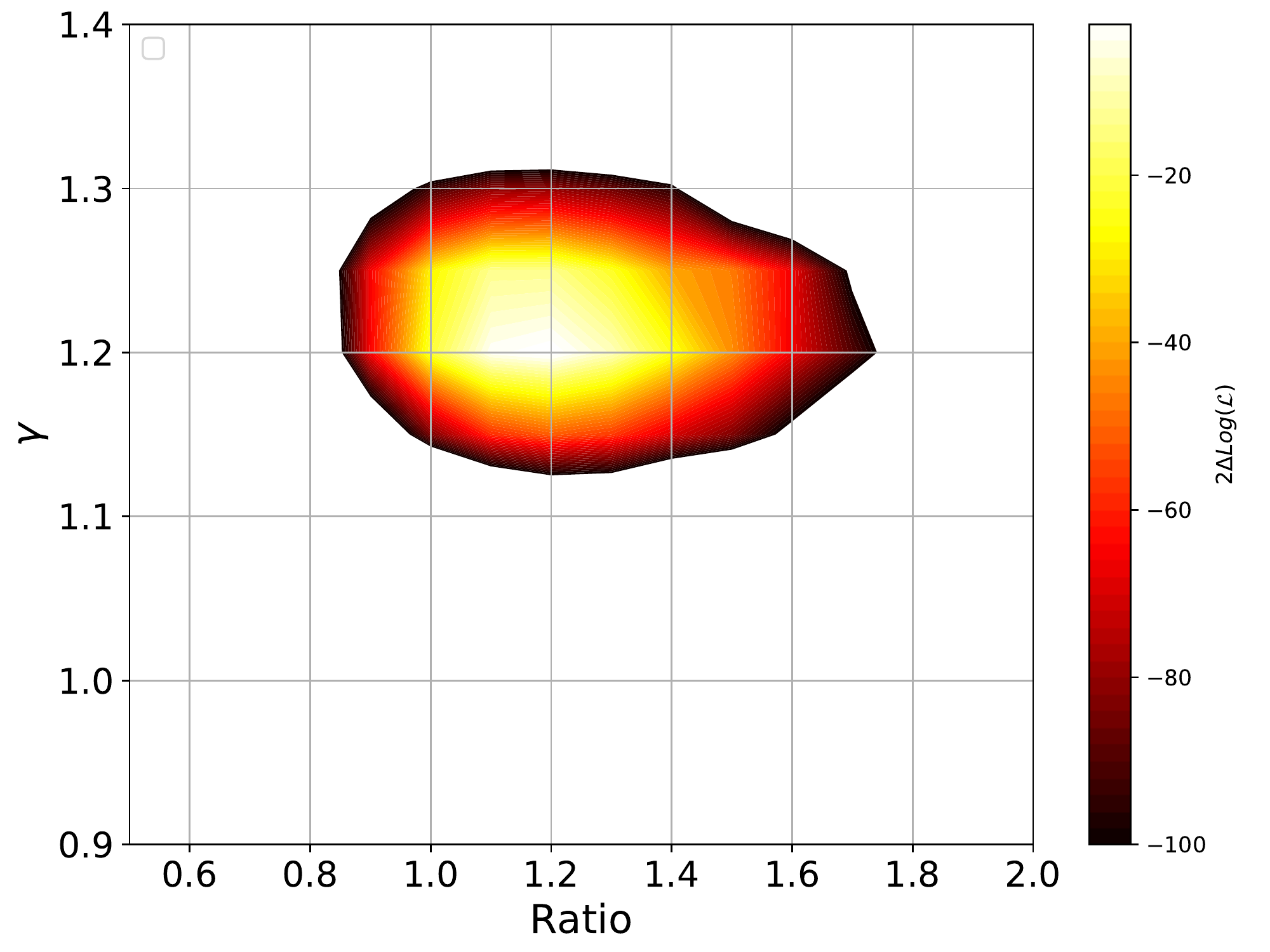}
\includegraphics[width=0.49\textwidth]{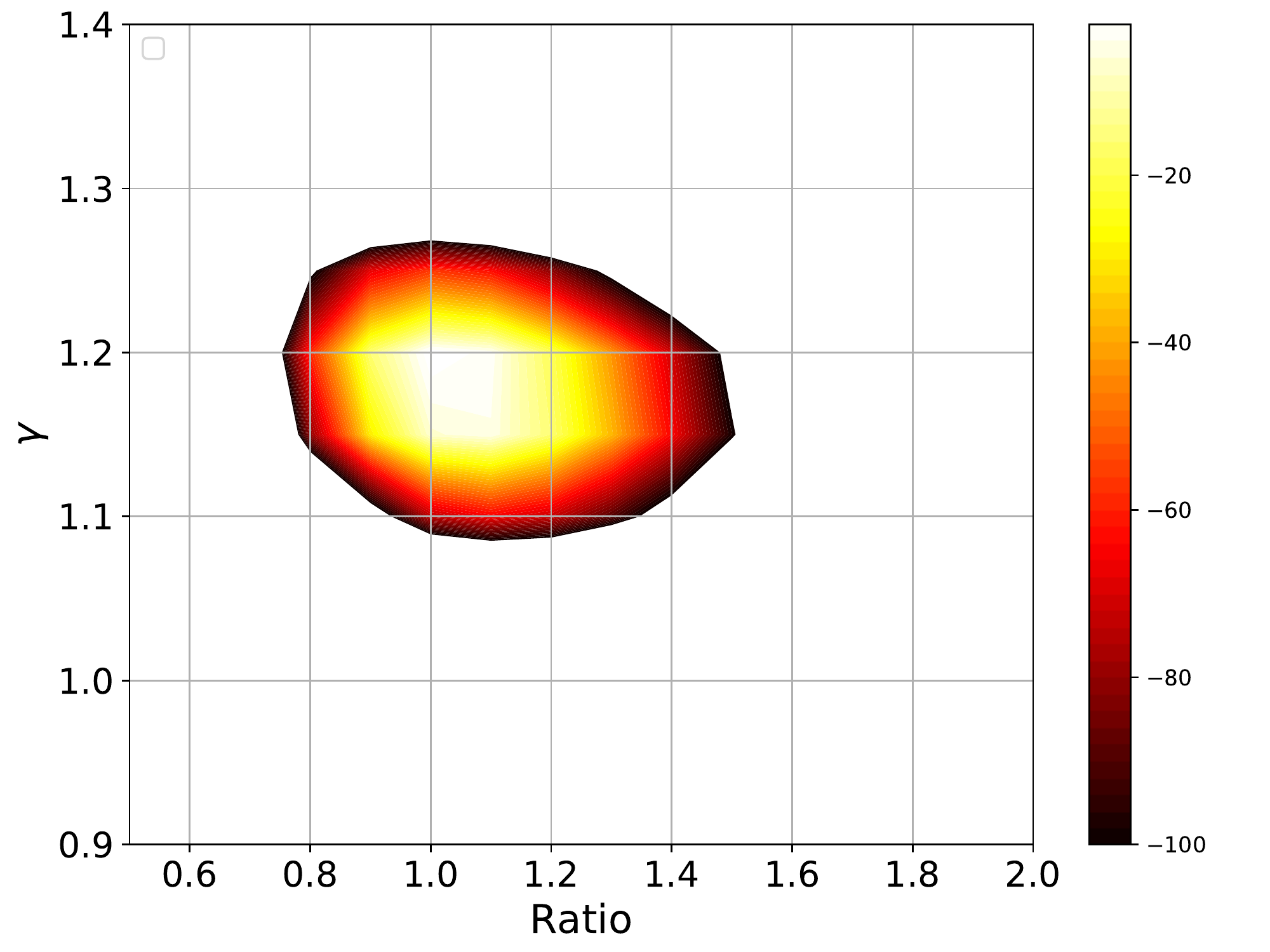}
\caption{Value of the likelihood as a function of the parameters {\it ratio} and $\gamma$ $\rm{Log}(\mathcal{L})(ratio,\gamma)$ found with the {\tt Baseline} (top left), {\tt SNR}  (top right), {\tt Yusifov} (bottom left) and {\tt SL ext} (bottom right) IEM for the energy range between $1-10$ GeV.}
\label{fig:sph}
\end{figure*}

\begin{table*}
\begin{center}
\begin{tabular}{|c|c|c|c|c|c|c|c|c|c|c|}
\hline
$\rm{Log}(\mathcal{L})-\rm{Log}(\mathcal{L}_{\rm{DM}})$   &   Baseline &  ICS combined  &   OB stars & Pulsars& SL ext. & SNR  &   Yusifov \\ 
\hline
BB  &  -1139	& -1192	& -797	& -1434	& -543	& -826	& -1043 \\
DM+NB &  +179	& +217	& +38	& +261	& +84	& +135	& +205 \\
BB+NB  &  +55(-124)	& +21(-196)	& -34(-72)	& +36(-225)	& -51(-135)	& +15(-120)	& +9(-196) \\
\hline
\end{tabular}
\caption{This table represents the difference of $\rm{Log}(\mathcal{L})$ obtained in the case of boxy bulge (BB), DM plus nuclear bulge (DM+NB) or boxy bulge and nuclear bulge (BB+NB) with respect to the case where we use only the DM template (DM). A positive number implies that the fit improves with respect to the case of the use of the DM template. In the last row we also report in parenthesis the difference of the likelihood of the boxy bulge and nuclear bulge with respect to DM plus nuclear bulge ($\rm{Log}(\mathcal{L_{\rm{BB+NB}}})-\rm{Log}(\mathcal{L}_{\rm{DM+NB}})$). In this case negative values imply that the fit with the boxy bulge and nuclear bulge is worse than the case with DM and the nuclear bulge.}
\label{tab:like}
\end{center}
\end{table*}

The last two GCE characteristics we investigate are the position and sphericity.
We still consider an energy range between $1-10$ GeV.

In order to find the position of the GCE we move the center of the DM template from $(l,b)=(0^{\circ},0^{\circ})$ as assumed before, to different locations around the Galactic center.
We choose a square of side $0.8^{\circ}$ and a grid of width $0.1^{\circ}$. 
We do not choose a finer grid because of the {\it Fermi}-LAT angular resolution.

In the analysis we use the same background model (i.e., sources and IEM components) and we only move the position of the DM template center.
For each DM template position we make a fit as explained in Sec.~\ref{sec:analysistec} and we save the value of $\rm{Log}(\mathcal{L})$. 
This analysis provides the likelihood as a function of the DM template position (longitude $l$ and latitude $b$) $\rm{Log}(\mathcal{L})(l,b)$.
The best-fit position is at about $l=-0.1^{\circ},b=-0.0^{\circ}$ for the {\tt Baseline} IEM.
The best fit position, considering all the IEMs and analysis setup, changes between $l=[-0.3^{\circ},0.0^{\circ}]$ and $b=[-0.1^{\circ},0.0^{\circ}]$.
Considering that the grid resolution is $0.1^{\circ}$ and that the best-fit position changes due to the IEMs and analysis setup more than the grid size, we conclude that the GCE position is compatible with the dynamical center, of the Milky Way that is considered to be Sagittarius A$^{\star}$, whose position is $(l,b) = (359.94,-0.05)$.
We remind that Sagittarius A$^{\star}$ is included in our background model with the source 4FGL J1745.6-2859.
We show in Fig.~\ref{fig:pos} the $\rm{Log}(\mathcal{L})$ as a function of the coordinates of the DM template obtained with the {\tt Baseline} model.
We also run this analysis for the other IEMs used in Fig.~\ref{fig:spectrumcases}, finding similar results for all of them. 
Refs.~\cite{Daylan:2014rsa,Linden:2016rcf} performed a similar analysis with only one IEM and found that the position is compatible with the Galactic center. 
On the other hand, Ref.~\cite{TheFermi-LAT:2017vmf} analyzed the position at different latitudes and found that the best fit for $b\in[-1^{\circ},1^{\circ}]$ is at $l \sim 1^{\circ}$. Moreover, the grid used in Ref.~\cite{TheFermi-LAT:2017vmf} to make the likelihood analysis as a function of the position seems to be of the order of $0.5^{\circ}$, i.e., much wider than what we assume in our paper.
Finally, Ref.~\cite{Karwin:2016tsw} found a best-fit position for the GCE template at roughly $-0.4^{\circ}$ of longitude and $0^{\circ}$ in latitude. They perform a likelihood analysis placing the DM template at different locations with a grid of $0.2^{\circ}$ spacing. Their results are thus similar to ours considering the larger grid size.

\subsection{Sphericity}
\label{sec:sphericity}

The sphericity of the GCE is one of the main features that can help to disentangle among the different interpretations.
Indeed, if the GCE is due to DM we expect a spherically symmetric signal.
In case the GCE is due to bremsstrahlung or $\pi^0$ decay we should have a signal elongated on the Galactic plane since these processes trace the distribution of gas in the Galaxy.
Finally, in \cite{Macias:2016nev,Macias:2019omb} the authors claim that using a boxy template, tracing old stars in the Galactic bulge, plus a template that follows the nuclear bulge, they are able to find a much better fit with respect to a DM template. 
The preference for the boxy bulge over the DM template should be demonstrated by finding that the GCE is not spherically symmetric.
Very recently, some of the authors of \cite{Macias:2016nev} have generated in \cite{Coleman:2019kax} a new model for the boxy bulge template that better models the stellar bulge and provides a better fit to the GCE. The authors of \cite{Bartels:2017vsx} reach similar conclusions for the preference of fitting the GCE with a template that follows the stellar bulge.

In order to study the GCE sphericity and investigate which of the above interpretations is more compatible with the GCE, we modify the DM template by introducing an ellipsoid to model its spatial distribution. In particular we introduce a parameter, called {\it ratio}, that parametrizes the ellipticity.
This is defined as the ratio between the axis along $l=(180^{\circ},0^{\circ})$ and $b=0^{\circ}$ and the axis on the perpendicular direction, i.e., defined along $l=0^{\circ}$ and $b=\pm90^{\circ}$. 
Therefore, an ellipsoid with a ratio larger than one implies that the template is elongated along the Galactic plane.
The boxy bulge model used in \cite{Macias:2016nev,Macias:2019omb,Coleman:2019kax} is very similar to an ellipsoid with ratio equal to 1.8.

We generate templates with different {\it ratio} values between 0.5 to 2.4 and for $\gamma$ from 0.8 to 1.7.
For each ellipsoid template, defined for a given {\it ratio} and $\gamma$ values, we perform a fit to the data. 
We show in Fig.~\ref{fig:sph} the contour plots for the likelihood as a function of ratio and $\gamma$ ($\rm{Log}(\mathcal{L})(ratio,\gamma)$) that we obtain when running the analysis with the {\tt Baseline}, {\tt SNR}, {\tt Yusifov} and {\tt SL ext} IEMs. 
With all the other IEMs and selections of the data and analysis techniques we reach similar conclusions.
We obtain best-fit values for the parameters of ratio $=0.8-1.2$ and $\gamma = 1.15-1.25$ meaning that there is not a clear preference for a GCE that is not spherically symmetric.
More importantly, the value of the ratio compatible with the boxy bulge used in \cite{Coleman:2019kax} is roughly 1.8.
This is very far from the maximum value we find using all the IEMs, which is 1.2.
Therefore, a spherically symmetric morphology seems to be preferred over a non-spherical template such as the boxy bulge model used in \cite{Macias:2016nev,Macias:2019omb,Coleman:2019kax}.
Our results for the ellipticity are consistent with the ones reported in \cite{Calore:2014xka,Daylan:2014rsa}.

However, the exercise done before does not provide any evidence that the spherically symmetric DM template performs better than the nuclear and boxy bulge templates. In fact, these latter are not exactly elliptical and the angular profile does not follow a NFW.
In order to test directly the interpretation published in Refs.~\cite{Macias:2016nev,Macias:2019omb,Coleman:2019kax}, we perform a fit in the energy range between $1-10$ GeV considering a DM template or the boxy bulge as reported in \cite{Coleman:2019kax}.
We also test in both cases the addition of the nuclear bulge template reported in \cite{Macias:2016nev}.
We use the same sources and IEM components for the DM and stellar bulge scenarios so the likelihood value of the fit is a direct information of which of the two scenarios fit better the GCE.
We generate the DM template using the best fit values for $\gamma$ found in Sec.~\ref{sec:spacial} with the fit to the GCE surface brightness data.

In Tab.~\ref{tab:like} we report the difference between the $\rm{Log}(\mathcal{L})$ obtained with the DM template (labelled as DM) and the following cases: with the boxy bulge (BB), DM plus the nuclear bulge (DD+NB) and the boxy bulge plus the nuclear bulge (BB+NB).
We show the results for different IEMs. For the models not reported in the table we find similar results.
The first thing we notice is that in the DM case the addition of the nuclear bulge template improves the fit by roughly $\Delta \rm{Log}(\mathcal{L})$ of $100-200$.
On the other hand, in the case of the boxy bulge the improvement is much larger $\sim500-1200$.
Finally, the case with DM only is preferred over the boxy bulge case for all the IEMs roughly for $\Delta \rm{Log}(\mathcal{L}) = 500-1200$.
The same is also true for the case DM+NB and BB+NB but with much smaller $\Delta \rm{Log}(\mathcal{L})$ values.
To conclude even if the addition of the nuclear bulge improves significantly the fit in the stellar bulge scenario, the DM interpretation still fits better the GCE for all IEMs considered in the analysis.
We have reported a similar test in Sec.~\ref{sec:spectrum} by performing a fit to the ROI using the {\tt IC bulge} model  after having deleted from it the DM template.
In this case, we were fitting the GCE with the inverse Compton emission from electrons and positrons injected from the Galactic bulge using the model in \cite{TheFermi-LAT:2017vmf}. The spatial template we used is reported in Fig.~\ref{fig:fluxmap} for a $\gamma$-ray energy of 1 GeV. 
This model is more complex than the one reported in Ref.~\cite{Macias:2016nev} because it is energy dependent since it is calculated fully solving the propagation equation of electrons and positrons in the Galactic bulge environment.
We tested different vertical sizes of the Galactic halo and for all the cases the fit with this template was worse than the DM template by $\Delta \rm{Log}(\mathcal{L}) \sim 230-250$ depending on the value for $z$. So the Galactic bulge model published in \cite{TheFermi-LAT:2017vmf} performs better than the one in Ref.~\cite{Macias:2016nev}. 

The different results we obtain with respect to Refs.~\cite{Macias:2016nev,Macias:2019omb,Coleman:2019kax} are probably due to the different assumptions we make for the background components. 
Probably, the most relevant ones are related to the choices for the IEM. Refs.~\cite{Macias:2016nev,Macias:2019omb,Coleman:2019kax} use interstellar gas components divided in rings (four each for the HI and H2 related emission) and include two dust correction templates too. These components are modeled with an energy independent morphology. The ICS emission is divided into 6 rings and the low-latitude component of the {\it Fermi} bubbles is also different with respect to the one we use. Finally, they include sources from the 2FIG catalog taken from Ref.~\cite{Fermi-LAT:2017yoi} while we use the 4FGL catalog.
There are other differences that are expected to provide a mild effect in the results such as selecting 7 years of LAT data instead of 11 as we do, using a smaller ROI of $30^{\circ}\times30^{\circ}$ and not using the weighted likelihood technique.
We will work on a follow-up analysis dedicated entirely to test the nuclear and boxy bulge models used in \cite{Macias:2016nev,Macias:2019omb,Coleman:2019kax} and on the investigation of the effect that the choice of different IEMs has on the preference of the bulge model over the DM scenario.

\section{Conclusions}
\label{sec: conclusions}
We analyze 11 years of {\it Fermi}-LAT data detected from the Galactic center region and we provide updated measurements of the characteristics of the GCE.
We utilize the state of the art IEMs created in \cite{TheFermi-LAT:2017vmf} for different source distributions, gas maps and inverse Compton configurations and test different selections of the data and analysis techniques.
Employing this strategy, we derive for the first time the systematics on several GCE properties due to the choice of several IEMs, data selections and analysis techniques.

First, we measure the energy spectrum which is peaked at a few GeV and is compatible with a log-parabola with spectral index of $-2.0$ and curvature index of 0.27.
We also find that the GCE SED normalization systematics are roughly a factor of 50\% at 2 GeV.
Our results are compatible with previous papers such as \cite{Calore:2014xka,TheFermi-LAT:2017vmf}.
We also verify that by including the emission of $\gamma$ rays produced by cosmic rays injected from the Galactic center in the IEM, the GCE spectrum is significantly modified, however, it is still significantly detected.

We investigate the spatial morphology of the GCE using a model independent technique based on an analysis using concentric and uniform annuli. 
Fitting the annuli to the data we find the surface brightness of the GCE and we verify that it is compatible with a DM template generated for a generalized NFW with $\gamma=1.1-1.2$ and an average value calculated among all the tested cases of 1.25.
The GCE extends with a significant flux up to roughly $12^{\circ}$ and its spatial morphology is well determined in the energy range between $0.6-30$ GeV.
We also use a model dependent technique where we use directly a DM template created with a generalized NFW and we confirm the previous results for the best-fit values of $\gamma$.

We also study the GCE taking a DM template and dividing it into quadrants. We derive the SED and the spatial distribution of each quadrant.
The SED is similar for each quadrant and well fit with the same SED found for the GCE modeled with a unique DM template.
The spatial distribution is also well fit with a DM template with $\gamma=1.1-1.3$, where the first quadrant (defined for $(l,b)>0$), has slightly larger values of $\gamma$.

We determine the best-fit position of the GCE centroid which is found to be $l=[-0.3^{\circ},0.0^{\circ}]$ and $b=[-0.1^{\circ},0.0^{\circ}]$ using the different IEMs and analysis setup of our analysis. Considering the variation in its value and the resolution of the grid we used in the analysis, which is $0.1^{\circ}$, the GCE position is thus compatible with the dynamical Milky Way center Sagittarius A$^{\star}$.

Finally, we demonstrate that the GCE is compatible with a spherically symmetric template considering the systematic uncertainties associated with the choice of the IEM.
Indeed, if we use a DM template modeled with an ellipsoid, we find that the ratio of the horizontal (i.e., along the Galactic plane) and vertical axis is between $0.8-1.2$.
Our results for the ellipticity are consistent with the ones reported in \cite{Calore:2014xka,Daylan:2014rsa}.
The best-fit region we find for the DM spatial morphology is not compatible with the one used in \cite{Macias:2016nev,Coleman:2019kax} for the boxy bulge which roughly similar to an ellipsoid with a ratio $\sim 1.8$. 
We tested this also with a fit to the data finding that the DM template provides a much better log-likelihood value with respect to the boxy bulge template derived in \cite{Macias:2016nev,Coleman:2019kax} or the model reported in \cite{TheFermi-LAT:2017vmf} that we label as {\tt IC bulge} and calculated for $\gamma$ rays produced for inverse Compton from electrons and positrons injected from the Galactic stellar bulge.
Our results differ from the ones in Refs.~\cite{Macias:2016nev,Macias:2019omb,Coleman:2019kax} where the boxy and nuclear bulge templates are preferred over the DM scanario. The difference is very likely due to different background components used in the analysis.

This paper thus provides evidence that the GCE SED has the following properties:
\begin{itemize}
\item {\bf DM SED}: the spectrum has a bump at around 2 GeV and is well described with a log-parabola with spectral index of $-2.0$ and curvature index of 0.27.
\item {\bf Spatial morphology}: the GCE is well modeled with a DM template parametrized with a generalized NFW density profile with $\gamma=1.1-1.2$. 
\item {\bf Energy evolution of spatial shape}: 
The energy evolution in the energy range $0.6-30$ GeV is constrained to be less than $10\%$ of the $\gamma$ value from its average value that is 1.25.
\item {\bf Position}: The GCE is centered at $l=[-0.3^{\circ},0.0^{\circ}]$ and $b=[-0.1^{\circ},0.0^{\circ}]$ and thus compatible with the dynamical Milky Way center.
\item {\bf Sphericity}: Considering an ellipsoid, the ratio between the horizontal, aligned along the Galactic plane, and vertical axis is between $0.8-1.2$, so the GCE is roughly spherically symmetric.
\end{itemize} 

We will present specific interpretations of the GCE in a companion paper. 
However, in Ref.~\cite{Dimaurosim} we have studied the properties of the GCE in case it is produced by the different mechanisms discussed in the introduction.
Therefore, we can already try to draw some conclusions.
The GCE from cosmic-ray protons would not be spherically symmetric but rather significantly elongated on the Galactic plane (see left panel of Fig.~\ref{fig:fluxmap}). 
The signal produced from cosmic-ray electrons would have a spatial extension that evolves with the energy and a spatial morphology that is not spherically symmetric  (see right panel of Fig.~\ref{fig:fluxmap}).
Indeed, we have found in Ref.~\cite{Dimaurosim} that the value of $\gamma$ would change by a factor of 2 in the energy range $1-100$ GeV and in this paper that the surface brightness changes shape significantly between 0.7 GeV and 20 GeV (see Fig.~\ref{fig:SB}). 
Finally, a millisecond pulsar population generated by the old stars present in the Galactic bulge would create a GCE that is compatible with all the properties we measured in this paper except for the morphology that, if modeled with the templates introduced in \cite{Macias:2016nev,Coleman:2019kax}, would not be spherically symmetric.
Indeed, a DM template is a better fit to the GCE than the boxy bulge scenario in \cite{Macias:2016nev,Coleman:2019kax}.
A millisecond pulsar population that is spherically symmetric would result in a fit that is as good as the one of DM.
Finally, a DM signal would be perfectly compatible with the GCE properties. However, the annihilation cross section needed to fit the GCE flux is constrained by the non-detection of $\gamma$-rays from Milky Way dwarf spheroidal galaxies (see, e.g., \cite{Hoof:2018hyn}).
Indeed, the GCE SED is roughly compatible with DM particles with a mass of 50 GeV and annihilating into a $b\bar{b}$ channel with an annihilation cross section of about $\langle \sigma v \rangle \sim 3-4  \times 10^{-26}$ cm$^3$/s\footnote{To obtain this result we have simply rescaled the results in \cite{Calore:2014xka}}. The upper limits found from Milky Way dwarf spheroidal galaxies are for the same DM candidate in the range $\langle  \sigma v \rangle < 2-3   \times 10^{-26}$ cm$^3$/2.

In order to further improve our knowledge of the GCE characteristics, a significant improvement in the modeling of the IEM is necessary.
Indeed, as we have demonstrated in this paper and in Ref.~\cite{Dimaurosim}, this component is the main source of systematic in the analysis of this complicated region of the sky. New gas maps, interstellar radiation field data and refined source distribution models need to be derived in order to achieve this goal.

\begin{acknowledgments}
The author thanks Michela Negro, Dmitry Malyshev, Chris Karwin for insightful discussions.

The author's research is supported by Fellini - Fellowship for Innovation at INFN, funded by the European Union's Horizon 2020 research programme under the Marie Sklodowska-Curie Cofund Action, grant agreement no.~754496.
He acknowledges support by the NASA Fermi Guest Investigator Program Cycle 12 through the Fermi Program N. 121119 (P.I.~MDM). 

The {\it Fermi} LAT Collaboration acknowledges generous ongoing support from a number of agencies and institutes that have supported both the development and the operation of the LAT as well as scientific data analysis. These include the National Aeronautics and Space Administration and the Department of Energy in the United States, the Commissariat\'a l'Energie Atomique and the Centre National de la Recherche Scientifique / Institut National de Physique Nucl\'eaire et de Physique des Particules in France, the Agenzia Spaziale Italiana and the Istituto Nazionale di Fisica Nucleare in Italy, the Ministry of Education, Culture, Sports, Science and Technology (MEXT), High Energy Accelerator Research Organization (KEK) and Japan Aerospace Exploration Agency (JAXA) in Japan, and the K. A. Wallenberg Foundation, the Swedish Research Council and the Swedish National Space Board in Sweden.
Additional support for science analysis during the operations phase is gratefully acknowledged from the Istituto Nazionale di Astrofisica in Italy and the Centre National d'Etudes Spatiales in France. This work performed in part under DOE Contract DE- AC02-76SF00515.
\end{acknowledgments}

\bibliography{paper}

\end{document}